\documentstyle[aps,epsf,amssymb,preprint]{revtex}

\newcommand{\di}{\mbox{$i\!\!\not\!\!D$}}
\newcommand{\nub}{\mbox{$\overline{\nu}$}}

\newcommand{\ssi}{\mbox{$\langle\overline{\psi}\psi\rangle$}}
\newcommand{\nni}{\mbox{$n_{\overline{I}}$}}

\begin{document}
\title{Chiral symmetry breaking and instantons \\
in both quenched and full QCD}
\author{U. Sharan  and  M. Teper}
\address{Theoretical Physics, University of Oxford, 1 Keble Road,
Oxford, OX1 3NP, U.K.}
\maketitle

\begin{abstract}
We investigate the contribution that instantons make to 
the QCD chiral condensate for $N_f=$ 0, 1 and 2 quark flavours.
We use a simplified model: the instantons
are a (weighted) gas, the fermionic integrations are 
restricted to the subspace spanned by the would-be zero-modes 
and only the dimensional and chiral features of the Dirac
operator are retained. Previous work along these lines
found a power-like divergence of the Dirac spectral density 
at zero eigenvalue. By changing the various components of the 
model, we show that a power divergence appears to be a generic feature 
of instanton mixing for both $N_f=0$ and $N_f\not=0$. In the 
latter case the divergence disappears as $m \to 0$ in such a way
that $\ssi$ remains finite. We find that the exponent of the power
decreases for larger instanton densities, becoming negligible
for very dense instanton `gases'. We reproduce the expected 
$m$-dependence of the topological susceptibility, we investigate 
the space-time structure of the eigenfunctions, as a function 
of the eigenvalue, and we calculate the $\eta^{\prime}$ mass. 
In summary: instantons appear to provide a natural mechanism for the 
spontaneous breaking of chiral symmetry in QCD, but they typically 
produce a power-like divergence in the Dirac spectral density. 
This divergence is non-standard but `harmless' in the case of 
full QCD, and implies a pathology in the case of quenched QCD.

\end{abstract}

\section{INTRODUCTION}
\label{section-intro}

The spontaneous breaking of chiral symmetry, and the dynamics that
drives it, remains a major unsolved problem in QCD.  However it has
long been recognised \cite{CDG-Carlitz,Diak,Shuryak} that instantons
provide a potential mechanism for this symmetry breaking. This can be
simply motivated using the well-known Banks-Casher formula
\cite{bnkcsh} and the Atiyah-Singer Index theorem \cite{atsi}.  The
Banks-Casher formula relates the value of the chiral condensate to the
density of modes of the Dirac operator, $\nub(\lambda)$, at zero
eigenvalue:
\begin{equation}
\label{eq:B-C}
\ssi = \pi\nub(0) ,
\end{equation}
while the Atiyah-Singer Index theorem tells us that the eigenvalue
spectrum of the Dirac operator will contain
\begin{equation}
\label{eq:atsi}
Q[A] = n_{-} - n_{+}
\end{equation}
exact zero modes, where $Q[A]$ is the winding number of the gauge
field configuration and $n_{\pm}$ are the number of zero eigenmodes
with positive/negative chirality. So if we neglect interactions
amongst (anti)instantons we get an exact zero-mode for each
topological charge in the gauge field.  This contributes a term
$\propto \delta(\lambda)$ to $\nub(\lambda)$.  That is to say, the
eigenvalues are clustered about zero.  This is in contrast to a purely
perturbative approach where, if we neglect interactions, the
perturbative vacuum becomes free and the eigenvalue spectrum grows as
$\lambda^3$. That is to say, the eigenvalues are far from zero.
Introducing interactions will shift both these distributions in ways
that are not easy to predict once the interactions are
strong. Nonetheless it is plausible that if what one wants is a non-zero
density of modes near $\lambda = 0$ then instantons provide a much
more promising starting point than the classical vacuum.

An old approach to obtaining chiral symmetry breaking from instantons
is as a self-consistent solution to semi-classical gap equations (see,
for example, \cite{mcdougall}).  An alternative approach is to perform
a direct calculation of $\ssi$ in some semi-classical approximation
(see, for example, \cite{Diak,Shuryak,ndmt}). Both approaches make
drastic approximations, but the latter has the attraction that these
can be easily made explicit. In this paper we develop upon the
calculations which were briefly summarised in \cite{ndmt}. We
now briefly outline our general approach.

To begin, we must make more precise what it might mean for chiral
symmetry breaking to be ``due to instantons''.  First, it is not at
all obvious that the topological charge of a typical gauge field must
be in a form that resembles an ensemble of instantons. We will,
however, assume that it is.  That is to say, we assume that it can be
decomposed into localised lumps of unit charge, $Q=\pm 1$, which can
be characterised by the usual instanton collective coordinates. We
shall refer to these lumps as `instantons' even though they will
normally be far from the minima of the action. There is some
evidence from lattice simulations for this assumption (see, for
example, \cite{DSMT,qlat-other}).  A second problem is that even if
the topological charge can be decomposed into an ensemble of
`instantons', then the correlations between these topological charges
will be influenced by other fluctuations in the gauge fields; for
example, those that induce confinement. So when we calculate the Dirac
spectrum associated with the instantons (see \cite{usmt2} for an
example of such a calculation and \cite{negele98} for a review of
related work) this will inevitably contain the
influence of other dynamics. Thus, even if we find that we do have
chiral symmetry breaking, it will be unclear to what extent it is due
to the instantons {\it per se} and to what extent it is due to this
other dynamics.  The simplest way to finesse this ambiguity is to
ignore all the correlations and see whether an uncorrelated ensemble
of instantons will, by itself, drive chiral symmetry breaking. If so,
if we can show that the spontaneous breaking of chiral symmetry
is a generic consequence of the most general features of an instanton 
ensemble, then we can plausibly claim that it is indeed `due to instantons'.

Since we are interested in the implications of instantons that are
generic rather than due to some special details of the approximations
used, it makes sense to avoid 
details altogether, as far as possible, at least as a first step. 
Such a minimalist approach, that attempts to include only
the essentials of the physics, was developed in \cite{ndmt,ndmt2} and
we shall use what is almost the same model here. For the pure 
gauge theory (i.e. $N_f=0$ or quenched QCD) we shall take our
instantons to have random positions and we give them some fixed size
$\rho$. The latter embodies the anomalous breaking of scale
invariance, while the former encodes the long distance clustering
properties of the field theory. We leave the density of instantons as
a free parameter and we ignore the other collective coordinates. If we
think of a single topological charge, its Dirac spectrum will contain
a zero eigenvalue and other, non-zero, modes. The former is
independent of how deformed the instanton is; the latter depend on all
the details. So if the symmetry breaking is to be a generic feature of
the instanton ensemble, it should depend only on the zero modes.  We
therefore restrict the fermionic path integral in the background of a
particular gauge field to the subspace spanned by the zero-modes of
the instantons in that gauge field. That is to say, we are
investigating whether chiral symmetry breaking is driven by the
would-be topological zero modes.  While the existence of a zero mode
is independent of detail, its functional form is not. Since we do not
know what this form is (and there is no reason to think that the
classical form is at all relevant) we will try a number of different
functional forms and see what, if anything, is insensitive to the
particular choice. In a matrix representation of the Dirac operator,
the overall structure is determined by the fact that $Q=\pm 1$ zero
modes have opposite chirality and that $\gamma_5$ anti-commutes with
$\di$. We shall assume that this is all that is important and simplify
the calculation accordingly. We shall use the same approach for
full QCD (i.e. $N_f \not= 0$) except that we weight the instanton
ensembles with the appropriate power of the fermionic determinant.
This determinant will also be calculated in the basis of would-be
zero-modes as described above.

We remark that the above may be thought of as a `bottom-up' approach
to this whole question.  There is an alternative approach that one
may think of as `top-down'. Here one uses random matrix theory (for
recent reviews see \cite{RMT}) identifying the appropriate
universality class on the basis of the symmetries in the problem, and
then making corresponding predictions for the fermionic physics.  To
what extent the two approaches are in agreement, even for quantities
for which they both make predictions, is not clear to us, but this is
a question that needs to be addressed.

The conclusion of \cite{ndmt,ndmt2} was not only that instantons
generically produce a non-zero chiral condensate, but that in quenched
QCD this condensate diverges as $m \to 0$. This divergence is a
reflection of a corresponding power-like divergence,

\begin{equation}
\overline{\nu}(\lambda) = a + b/\lambda^{d},
\label{eqn-power-law}
\end{equation}
as $\lambda \to 0$, in the spectral density. The value of the exponent
was found \cite{ndmt,ndmt2} to be $d \simeq 1/2$.
If this is indeed the case, then it suggests a striking pathology
in quenched QCD. An essential first step in assessing this result is
to determine how robust it is against changes in the assumptions
employed.  In particular, the zero-modes were modeled, in
\cite{ndmt,ndmt2}, by simple hard-sphere wave-functions (and the
wavefunction overlaps were generally calculated using only the
space-time points at which the topological charges were located). In
the present paper we shall compare calculations with several different
functional forms (and we shall use the full overlap functions).  We
shall also perform a detailed study of finite-volume effects and of
how things depend on the instanton density or `packing fraction'
$f$. The latter has a natural definition for `hard sphere'
wavefunctions:

\begin{equation}
f \equiv {{\overline{N} \ \overline{V}_I}\over{V}}
\label{eqn-pack-frac}
\end{equation}
where $\overline{N}$ is the mean number of topological objects in
the gas, $\overline{V}_{I}$ is the mean volume of such an object and
$V$ is the volume of space-time.  (Recall that the volume of a
4-sphere of radius $\rho$ is ${V}_{I}=\pi^2\rho^4/2$.)  What is the
corresponding definition of $f$ for other wavefunctions is not
immediately obvious and is one of the questions we shall address
below. Our calculations will lead us to a somewhat more nuanced
conclusion than that drawn in \cite{ndmt}.  In particular we find that
while a divergence is indeed a characteristic product of the mixing of
the would-be zero-modes, the value of the exponent depends quite
strongly on the instanton density, becoming negligible for large
packing fractions.

In full QCD there should be no such pathology, and so it is of 
interest to ask what happens to this small-$\lambda$ divergence
as we make $N_f \not= 0$. We find that the divergence is still
there but that its coefficient $\to 0$ as the fermion mass 
$\to 0$ in (apparently) such a way that the chiral condensate 
has the conventional smooth chiral limit. Thus the divergence is
`harmless' although the Dirac spectrum is unconventional
and the usual Banks-Casher\cite{bnkcsh} relation, given in 
equation~\ref{eq:B-C}, need no longer be valid.

In the next section we provide a brief discussion of instantons and
chiral symmetry breaking which will also serve to introduce our
notation. We then introduce our minimalist model for an ensemble of
instantons, their associated zero modes and our matrix representation
for the Dirac operator.  We point to some calculational difficulties
and how we deal with these by the choice of an appropriate space-time
manifold. We then turn to the results of our calculations for
quenched QCD. Our main
result is that the Dirac spectrum diverges at small eigenvalues. How
this result depends on details, both unphysical -- such as the
functions chosen for the zero-modes and the space-time manifold in
which they exist -- and physical -- such as the instanton density and
the physical volume -- is investigated in detail. We also investigate
whether the eigenfunction of the Dirac operator changes character as
the corresponding eigenvalue decreases. We find that indeed it does:
as $\lambda \to 0$ the eigenfunction becomes much more non-local.
Having acquired some confidence that the qualitative features
of our calculations are independent of the specific model assumptions,
we then proceed to the (computationally) more difficult case
of full QCD. Here we ask whether instantons spontaneously
break chiral symmetry, what happens to the $\lambda \to 0$ 
divergence in the Dirac spectral density, and whether our
model can correctly reproduce the $m$-dependence of the topological 
susceptibilty that is predicited by the anomalous Ward identities.
We also calculate the $\eta^{\prime}$ mass and confirm that
it is not a Goldstone boson.

Some of our early results were summarised in \cite{usmt}.  A
calculation of the Dirac spectrum using the cooled instanton ensembles
obtained in the SU(3) lattice gauge theory calculation of \cite{DSMT} is
to be found in \cite{usmt2}.  Although the results differ, in the fine
detail, from what one obtains with a randomly placed ensemble of
instantons, the main features of spontaneous symmetry breaking and a
divergent quenched chiral condensate appear to be present.

\section{CHIRAL SYMMETRY BREAKING AND TOPOLOGY}
\label{section-chiral}

The usual order parameter for chiral symmetry (breaking) is 
the chiral condensate in the limit of infinite volume and
zero mass: $\ssi = \lim_{m\to 0} \lim_{V\to \infty}\ssi_{m,V}$.
We can express $\ssi_{m,V}$ in terms of the normalised Dirac 
spectral density, $\nub(\lambda,m)$, as

\begin{equation}
\ssi_{m,V} = \frac{\overline{N}_{Z}}{mV} +
\int_{0}^{\infty}\frac{2m\nub(\lambda,m)}{\lambda^{2} + m^{2}} 
d\lambda.
\label{eqn-chico}
\end{equation}
Here we have chosen to separate from $\nub$ the $\delta$-function
contribution of the exact zero modes, whose mean number is
${\overline{N}_{Z}} = {\overline{|Q|}} \propto \surd V$.  
In the thermodynamic
limit this zero-mode contribution will clearly disappear. Thus one
might expect that at finite-$V$ the eigenvalue spectrum restricted to
$Q=0$ field configurations will have smaller finite-$V$ corrections
than if we allow $Q$ to vary. (As long as the total number of
topological charges is $N \gg 1$, so that we would have had
$\langle Q^2 \rangle \gg 1$ if we had allowed $Q$ to vary, this 
should not represent an unphysical bias on the calculation.)
In this expectation (to be verified later on) most of the calculations
in this paper will be for fields with $Q=0$. 

Note that if we work in a finite volume, then we would normally choose
it to be at least large enough to accommodate the lightest particle in
the theory, the Goldstone pion, i.e. $V^{\frac{1}{4}} \sim
1/m_{\pi}$.  If we insert $m^2_{\pi}{f^2_{\pi}} = {m\ssi}$ and crudely
estimate ${\overline{N}_{Z}} \simeq \surd \langle Q^2 \rangle =
\sqrt{\chi_t V}$, where $\chi_t$ is the topological susceptibility,
then we find that the zero-mode contribution in
equation~\ref{eqn-chico} is

\begin{equation}
\frac{\overline{N}_{Z}}{mV} \sim \frac{\chi^{\frac{1}{2}}_t}{f^2_{\pi}} \ssi
\sim 4  \ssi
\label{eqn-chico-zm}
\end{equation}
where we have used the standard values, $\chi_t \sim (180 MeV)^4$ and
$f_{\pi} \sim 93 MeV$. This demonstrates that the finite-volume
correction to the quenched QCD chiral condensate from the exact
zero-modes can be large even if one uses space-time volumes that might
appear to be adequate.

In full QCD the above calculation will differ in that the topological
susceptibility will vary with the quark mass: $\chi_t \simeq m
\ssi/n^2_f$ if we have spontaneous chiral symmetry breaking. Inserting
this into equation~\ref{eqn-chico-zm} we find that the zero-mode
contribution

\begin{equation}
\frac{\overline{N}_{Z}}{mV} 
\sim 
\frac{\sqrt{\chi_tV}}{mV}
\sim 
\frac{1}{(mV)^{\frac{1}{2}}}\frac{\ssi^{\frac{1}{2}}}{N_f}
\propto
m^{\frac{1}{2}}
\label{eqn-chico-zm-qcd}
\end{equation}
vanishes when we take the
chiral limit while maintaining $V$ large enough to accommodate the
Goldstone pion. Note that the case of a single flavour, $N_f = 1$, is
special. Now there will be no Goldstone pion because the $U_A(1)$
symmetry is anomalous: the would-be pion is a massive $\eta^\prime$.
So there is a finite mass-gap in the theory as $m \to 0$.  In such a
situation one would normally expect that a fixed finite volume, with
$V^{\frac{1}{4}} > 1/m_{\eta^\prime}$, would be large enough. However
we note that this would lead to a $\propto 1/\sqrt{m}$ divergence in
the chiral condensate.

One can ask what happens to the zero-mode contribution if one 
chooses to take the chiral limit in a fixed volume. In that
case we expect symmetry restoration for sufficiently small $m$,
and for smaller masses we may expect $\chi_t \propto m^{N_f}$.
Thus for $N_f \geq 2$ there will be no real $m\to 0$ divergence
in $\ssi_{m,V}$ at fixed $V$.
 
The above finite volume corrections, due to the $|Q|$ exact zero
modes, induce a trivial divergence in the 
$N_f = 0$ chiral condensate if we take
the $m \to 0$ chiral limit at fixed volume.  As remarked earlier, we
shall evade this divergence by usually working with $Q=0$ field
configurations. However, as described in the introduction, there is
evidence \cite{ndmt,ndmt2,usmt} for a much less trivial divergence;
one which survives the thermodynamic limit and which appears as a
divergence in the spectral density of the form in
equation~\ref{eqn-power-law}.  One of the main aims of this paper is
to determine whether this divergence is really there, 
in both quenched and full QCD,
or whether it is an artefact of the approximations used in the
calculations of \cite{ndmt,ndmt2,usmt}.

There is a simple argument \cite{ndmt,ndmt2}, briefly described in
\cite{ndmt}, that provides an apparently natural explanation for such
a divergence. It goes as follows.

In a finite volume $V$ one expects to lose spontaneous symmetry
breaking once the explicit symmetry breaking term is small enough
i.e. once $mV$ becomes small. That is to say, we expect a suppression
in the finite volume spectral density for $|\lambda| < O(1/V)$. Let us
assume, for simplicity, that the only finite-$V$ correction affecting
the spectral density of the non-zero modes consists of a cut-off for
$|\lambda| < c/V$.  Then the only other finite-$V$ correction concerns
the exact zero-modes. If we increase the volume from $V$ to $V^\prime$
then their number increases from 
$n_z \sim \sqrt{\langle Q^2 \rangle} \sim \sqrt{\chi_t V}$ to $\sim
\sqrt{\chi_t V^{\prime}}$.  Therefore, as a fraction of the total
number of modes,
their number will decrease from $\sim \sqrt{\chi_t/V}$ to $\sim
\sqrt{\chi_t/V^{\prime}}$. The integrated spectral density (including
the zero modes) does not change with volume so these two finite-$V$
corrections must match:

\begin{equation}
\int_{\frac{c}{V^{\prime}}}^{\frac{c}{V}} \nub(\lambda) d\lambda =
\sqrt{\frac{\chi_t}{V}} - \sqrt{\frac{\chi_t}{V^\prime}}.
\label{nz-match}
\end{equation}
where  $\nub(\lambda)$ is the spectral density for $V = \infty$.
It is easy to see that this relation implies that

\begin{equation}
\nub(\lambda) \stackrel{\lambda \to 0}\propto \lambda^{-\frac{1}{2}}
\end{equation}
and hence that $\ssi \propto m^{-\frac{1}{2}}$ in the thermodynamic,
chiral limit.

The important assumption in this argument is that as we increase the
volume, the fractional decrease in the exact zero-modes is matched by
an increase amongst the very smallest of the non-zero modes. We shall
refer to this as the ``eigenvalue replacement hypothesis'' and shall
investigate it in more detail later on in this paper. For now we
simply point out that intuitively it is not unreasonable. In a volume
$V$ we expect on the average $c_0\sqrt{V}$ exact zero modes. Now
consider this volume split into two equal halves. Considered in
isolation each half-volume would possess $\sim c_0\sqrt{V/2}$ exact
zero-modes.  So in joining these two half-volumes together we have
decreased the total number of zero modes from $\sim 2 \times
c_0\sqrt{V/2} = c_0\sqrt{2V}$ to $c_0\sqrt{V}$.  That means some of
the original zero-modes have mixed and split from zero. But if $V$ is
sufficiently large, one would expect the zero modes in the two halves
to be mostly far from each other so that the overlap will be very
small, and the eigenvalues will be split very little from zero. This
very qualitative argument should really be formulated using a number
of subvolumes for which one can use statistical arguments.  But it
suffices to give some motivation for what we have called the
eigenvalue replacement hypothesis.

The above argument would seem to be equally applicable to full QCD,
with one modification: now $\chi_t \propto m$ (as long as chiral 
symmetry is spontaneously broken). Inserting this into 
equation~\ref{nz-match} we find

\begin{equation}
\nub(\lambda) \stackrel{\lambda \to 0}\propto 
\Bigl({\frac{m}{\lambda}}\Bigr)^{\frac{1}{2}}.
\label{div-qcd}
\end{equation}
While this is an unconventional form for the spectral density,
it does not create any divergence in the chiral condensate:

\begin{eqnarray}
\ssi  
& = & 
\lim_{m \to 0}  \lim_{V \to \infty} \ssi_{m,V}\nonumber\\
& = & 
\lim_{m \to 0} 
\int_{0}^{\infty}\frac{2m\nub(\lambda,m)}{\lambda^{2} + m^{2}} 
d\lambda \nonumber\\
& = & 
\lim_{m \to 0} 
\int_{0}^{\infty}\frac{2m\{ a + b({{m}\over{\lambda}})^{\frac{1}{2}}\}}
{\lambda^{2} + m^{2}} d\lambda 
\nonumber\\
& = & 
\int_{0}^{\infty}\frac{2\{ a + {{b}\over{y^{\frac{1}{2}}}}\}}{y^2 + 1} 
dy \nonumber\\
& = & 
\pi a + 2b\int_{0}^{\infty}\frac{dy}{y^{\frac{1}{2}}(y^2 + 1)} .
\end{eqnarray}
Here we have used a spectrum 
$\nub(\lambda) = a + b(m/\lambda)^{\frac{1}{2}}$ and have performed
a change of variable to $y = \lambda/m$. We see that the result
for $\ssi$ is finite. However the naive application of the
Banks-Casher formula, which would have produced the result $a\pi$,
breaks down. So, although a spectral density of this kind does not
immediately lead to any pathological physics, it is clear that
it might undermine any argument which relies on the 
$\lambda \to 0$ and $m \to 0$ limits commuting with each other. 

The form we conjecture in equation~\ref{div-qcd} can also be
made plausible as follows. Since a mode of $\di$ will
not be (individually) suppressed by the  
$(\det\di)^{N_f} = \prod (\lambda^2_i + m^2)^{N_f}$ weighting
if $\lambda \ll m$, we might expect that the spectral 
density for full QCD will have the same form as in quenched
QCD for $\lambda \ll m$. If the quenched spectrum behaves as
$1/\lambda^{\frac{1}{2}}$ as $\lambda \to 0$ then the spectrum 
of full QCD should have the form 
$\nub(\lambda,m) = (\mu/\lambda)^{\frac{1}{2}}$ for
$\lambda \ll m$ where $\mu$ will have dimensions of mass, just
on dimensional grounds. The theory has two dimensionful parameters:
the quark mass $m$ and the pure gauge scale $\Lambda_{QCD}$. Since 
the term $(\mu/\lambda)^{\frac{1}{2}}$ should only become important 
(compared to the constant term say) for $\lambda \ll m$, it must
be that $\mu \propto m$. That is to say, the form in 
equation~\ref{div-qcd}. 

If we use the argument in the previous paragraph, then we can 
reverse our earlier argument in order to show that in QCD we should 
have $\langle Q^2 \rangle \propto mV$. This is of interest because
it is actually hard to find a simple explicit argument that gives 
this result. (The usual derivation using anomalous Ward identities
may be simple but is hardly explicit.) For example, if we take 
the dilute gas limit then each topological charge has its
separate fermionic weighting $\propto (\det\di[A_I])^{N_f}
\propto m^{N_f}$ that accompanies  the factor of $V$
(the charge can be located anywhere in the space-time volume). One 
immediately concludes that $\langle Q^2 \rangle \propto m^{N_f} V$.
Suppose instead we try the opposite extreme, where the only
important effect of the fermionic weighting is that the gauge field 
acquires a factor $\propto m^{N_f|Q|}$. If we had just a pure 
gauge weighting, then we would have $\langle Q^2 \rangle \propto V$,
and for $Q^2 \ll V$ the weighting will be 
almost independent of $Q$.
For such $Q$ the only non-trivial weighting is our assumed fermionic 
weighting  $\propto m^{N_f|Q|} = \exp\{-N_f|Q|\ln(1/m)\}$. It
is easy to see that this implies $\langle Q^2 \rangle \propto
m^{N_f}$. (Now we do not even have the correct factor of $V$.)
It is clear that the real situation is much more subtle. When we 
have chiral symmetry breaking the spectral density remains
finite all the way down to $\lambda = 0$. So the $|Q|$ exact
zero-modes may have been traded for modes that were arbitrarily close
to zero. It is this effect that presumably reduces the
naive $\langle Q^2 \rangle \propto m^{N_f}V$ dependence (which 
is correct when we have chiral symmetry restoration and no
non-trivial density of modes arbitrarily close to zero) to the 
weaker $\langle Q^2 \rangle \propto mV$ dependence. It will
be interesting to see whether we can reproduce this subtle trade-off
within the simple model that we shall be using in this paper.

\section{THE MODEL}
\label{section-model}

In this section we describe how we are going to calculate the
contribution of instantons to the Dirac spectral density.  This
requires making a number of radical simplifications.  Although we
shall try to give each of these simplifications some motivation, it is
important to perform some check that the final conclusions are robust
against changes in the assumptions being made.

We shall first describe the general framework of our calculation.  (We
refer the reader to \cite{usmt2} for a detailed discussion of
some aspects of the model.) We shall then describe the various
zero-mode wave functions that we use in our calculations.  Comparing
the results of these calculations will provide a check of the
robustness of our scheme.

\subsection{The general strategy}

We are interested in determining what are the generic physical effects
of instantons. Suppose we have a gauge field which contains $n_{I}$
instantons and $n_{\overline{I}}$ anti-instantons.  Suppose that the
fields of the instantons and anti-instantons, if in isolation, would
be $ A_{i}^{+}$ and $ A_{i}^{-}$ respectively. Using a linear addition
approximation one can write a field
\begin{equation}
A = \sum_{i}^{n_{\overline{I}}} A_{i}^{+} + \sum_{j}^{n_{I}} A_{j}^{-}.
\end{equation}
that represents the instanton content of the original gauge field.
(We suppose that each individual instanton field is in a suitable
singular gauge so as to be as local as possible.)  One could do this
for a realistic ensemble of (lattice) gauge fields, as was done in
\cite{usmt2}. However in that case there will be correlations amongst
the instantons that are partly determined by other, non-topological
fluctuations and if we are interested in the generic properties of
intantons {\it per se} then we do not want to include these. So the
fields we shall consider will be those due to a randomly positioned
ensemble of instantons. Such an ensemble clearly possesses the
long-distance clustering properties of the field theory. We shall give
the instantons a specific size; this embodies the anomalous
scale-breaking in the theory. The instanton size sets the length scale
of the theory and the volume of space-time can be expressed in those
units. Varying the average distance between instantons amounts to
varying the density, or packing fraction, as defined earlier.  A
slightly more elaborate approach would be to give the instantons a
distribution of sizes; in particular for small sizes the distribution
is calculable in perturbation theory.  For the results of such an
approach we refer the reader to \cite{usmt2}.

The above is plausible for quenched QCD, but breaks down for
full QCD where the correlations between topological charges will
become of increasingly long range as $m \to 0$. This can be 
incorporated by weighting the instanton ensemble by a
factor $\propto (\det\di)^{N_f}$. This will be described in
detail later on in the paper.

There are a total of $N[A] = n_{\overline{I}} + n_{I}$ objects in our
gauge field $A$ (in the volume $V$), and the winding number of $A$ is
$Q[A] = n_{I} - n_{\overline{I}}$. We know via the Atiyah-Singer Index
Theorem \cite{atsi} that the Dirac operator in this background
$\di[A]$ will have at least $Q[A]$ exact zero eigenvalues. If the
objects in the superposition were non-interacting then each would have
a corresponding zero eigenvalue,
\begin{equation}
\di[A_{i}^{\pm}]\psi_{i0}^{\pm} = 0,
\label{wbzero}
\end{equation}
again due to the Atiyah-Singer Index Theorem. We see from
equation~\ref{wbzero} that these zero modes are modes of definite
chirality, in fact $\gamma^{5}|\psi_{i0}^{\pm}\rangle =
\pm|\psi_{i0}^{\pm}\rangle$. Interactions between the objects however,
lead to a reduction of exact zero modes from $N[A]$ to only
$Q[A]$. The remaining $N[A] - Q[A]$ modes have split symmetrically
about zero \cite{Shuryak,Diak,usmt2}. We shall frequently refer to
these modes as the ``would-be zero modes''. The model, which is by no
means new, is based around the idea of constructing an explicit matrix
representation of the Dirac operator, using only the zero mode
wavefunctions $\{|\psi_{i0}^{+}\rangle, |\psi_{j0}^{-}\rangle\}$ as a
basis. Of course, the claim is not that these zero mode wavefunctions
in isolation form a basis (they certainly do not span), the claim is
that by constructing the Dirac operator on this subspace we can
understand that part of the Dirac spectrum which arises from instanton
interactions. Furthermore as these are the eigenvalues which generate
a spectral density around zero, we may be able to comment upon the
idea of instantons as the source of chiral symmetry breaking in
QCD. We drop the subscript zeroes on the would be zero mode
wavefunctions from now on, as these are the only wavefunctions which
we shall consider. It is easy to see that:

\begin{eqnarray}
\langle\psi_{i}^{+}|\di[A]|\psi_{j}^{+}\rangle & = & 0\nonumber\\
\langle\psi_{i}^{-}|\di[A]|\psi_{j}^{-}\rangle & = & 0\nonumber\\
\langle\psi_{i}^{+}|\di[A]|\psi_{j}^{-}\rangle & \doteq &
V_{ij} .
\end{eqnarray}
The first two identies follow from $\{\di,\gamma^{5}\}\ =\ 0$ and the
definite chirality of the zero mode wavefunctions. The function
$V_{ij}$ can in principle depend upon the size, position and relative
colour orientation of all the objects in the field. We shall ignore
the colour orientation of the objects for simplicity. And we shall
assume that $V_{ij}$ depends only on the objects $i$ and $j$. This is
an approximation that becomes increasingly hard to justify as the
instantons become denser. See \cite{usmt2} for a more detailed
discussion. We also note that:

\begin{eqnarray}
\langle\psi_{i}^{+}|\psi_{j}^{+}\rangle & = & U(x_{i}^{+},\rho_{i}^{+},x_{j}^{+},\rho_{j}^{+})\nonumber\\
\langle\psi_{i}^{-}|\psi_{j}^{-}\rangle & = & U(x_{i}^{-},\rho_{i}^{-},x_{j}^{-},\rho_{j}^{-})\nonumber\\
\langle\psi_{i}^{\pm}|\psi_{j}^{\mp}\rangle & = & 0 ,
\label{non-orth}
\end{eqnarray}
simply from the chiral properties of the zero mode wavefunctions. The
function $U$ will depend upon the size, position and relative colour
orientation (which we ignore) of the objects. We now use
equation~\ref{non-orth} to construct an orthonormal set of vectors
$\{|\widetilde{\psi}_{i}^{+}\rangle,\
|\widetilde{\psi}_{j}^{-}\rangle\}$ using Gram-Schmidt
orthonormalization:

\begin{eqnarray}
|\psi_{j}^{+}\rangle & = & R_{ij}|\widetilde{\psi}_{i}^{+}\rangle\quad 1 \leq
i \leq j \leq \nni\nonumber\\
|\psi_{j}^{-}\rangle & = & S_{ij}|\widetilde{\psi}_{i}^{-}\rangle\quad 1 \leq
i \leq j \leq n_{I} .
\end{eqnarray}
A representation of the Dirac operator (in this restricted subspace)
is given by:

\begin{equation}
\begin{array}{ccc}
\hspace{1.8cm}\overbrace{\hspace{2.7cm}}^{n_{\overline{I}}} & \hspace{0.0cm}\overbrace{\hspace{2.7cm}}^{n_{I}} & \\
\di \doteq \widetilde{D} = 
\left(
\begin{array}{l}
\langle\widetilde{\psi}_{i}^{+}|\di|\widetilde{\psi}_{j}^{+}\rangle = 0 \\
\langle\widetilde{\psi}_{i}^{-}|\di|\widetilde{\psi}_{j}^{+}\rangle = \widetilde{V}_{ij}^{\dagger}
\end{array}
\right. & 
\left.
\begin{array}{l}
\langle\widetilde{\psi}_{i}^{+}|\di|\widetilde{\psi}_{j}^{-}\rangle =
 \widetilde{V}_{ij}\\
\langle\widetilde{\psi}_{i}^{-}|\di|\widetilde{\psi}_{j}^{-}\rangle = 0
\end{array}
\right) &
\begin{array}{l}
\Big\}\ \nni\\ \Big\}\ n_{I}
\end{array}
\end{array}
\label{eq:dirm}
\end{equation}
where
\begin{eqnarray}
\widetilde{V} & = & (R^{-1})^{\dagger}VS^{-1}
\end{eqnarray}

We can therefore construct this matrix representation for any given
configuration of instantons providing we have been given the functions
$U(x_{i}^{\pm},\rho_{i}^{\pm},x_{j}^{\pm},\rho_{j}^{\pm})$ and
$V(x_{i}^{\pm},\rho_{i}^{\pm},x_{j}^{\mp},\rho_{j}^{\mp})$. Once we
have the matrix it is a simple matter to calculate its eigenvalues and
build up a spectral density using an ensemble of configurations. 

The function $U$ will depend on the functional form of the would-be
zero modes. What this form is, in a realistic gauge field, is not
known, apart from the fact that one expects it to be localised around
the centre of the instanton and to have an extent that is related to
the size of the instanton. (We remark that there is no particular
reason to favour the known semi-classical zero-mode solution.) We
assume that apart from the chirality, the spinorial degrees of freedom
are inessential.  We will therefore use wave-functions that are simple
real valued functions of position, with an index, $+$ or $-$, to label
the chirality. We shall use several simple trial functional forms, as
described in detail below.  Only those features of the resulting Dirac
spectrum that are independent of the wave-function form used, can be
considered to be generic features of instanton physics.

The function $V$ depends not only on the form of the would-be zero
modes but also on what we choose for $\di$. If all the other fields
have a negligible presence in the region where $\psi_{l}^{-}(x)$ and
$\psi_{k}^{+}(x)$ are non-zero, then we can write

\begin{eqnarray}
V_{kl} & = & \langle\psi_{k}^{+}|\di[A]|\psi_{l}^{-}\rangle\nonumber\\
& \approx & \langle\psi_{k}^{+}|\di[\sum_{i} A_{i}^{+} + \sum_{j}
A_{j}^{-}]|\psi_{l}^{-}\rangle\nonumber\\
& \approx & \langle\psi_{k}^{+}|\di[A_{k}^{+} +
A_{l}^{-}]|\psi_{l}^{-}\rangle\nonumber\\
& = & \langle\psi_{k}^{+}| \di[A_{k}^{+}] + \di[A_{l}^{-}] -
i\!\not\!\partial |\psi_{l}^{-}\rangle\nonumber\\
& = & \langle\psi_{k}^{+}|-i\!\not\!\partial|\psi_{l}^{-}\rangle
\label{eqn-D-to-d}
\end{eqnarray}
Here the first approximation is a consequence of the linear addition
ansatz as stated previously. The second approximation holds if the
objects are well separated from one another, however it does not hold
in general (especially for the high density gases to which we
sometimes apply our model). It is however necessary if we are to
progress beyond the ``dilute'' gas approximation using a model such as
ours. The last two equalities hold due to the fact that the states are
zero-modes: $\di[A_{l}^{\pm}]|\psi_{l}^{\pm}\rangle = 0$.  Of course
in any realistic situation this matrix element will depend in a
complicated way on all the other fields. We can only hope to elucidate
those properties that are independent of this unknown dependence. In
that case it makes sense to begin by simplifying this operator as
radically as possible; that is to say, we retain only its
dimensionality and its chiral properties. So, apart from anticommuting
with $\gamma_5$, we assume that the Dirac operator acts like a unit
operator multiplied by a scale factor $1/\rho$ where $\rho$ is a
characteristic length scale of the matrix element under
consideration. We choose to use $\rho \equiv \sqrt{\rho_i\rho_j}$ for
the matrix element $V_{ij}$ (so ignoring the scale provided by the
distance between the instanton and anti-instanton):

\begin{equation}
V_{ij} \simeq {1 \over{\sqrt{\rho_i\rho_j}}} \int d^4x
\psi_{j}^{-}(x)\psi_{i}^{+}(x) 
\label{eqn-D-I}
\end{equation}
While extreme, we note that this choice has the correct qualitative
behaviour in the dilute gas limit: all the eigenvalues go to zero.  Of
course in this limit the splitting from zero will depend on the
detailed form of the tails of the zero-modes. In the opposite limit,
where the density becomes large, our basis of zero-modes eventually
becomes complete and it is easy to see that our matrix $V$ will tend
to the unit matrix divided by the size $\rho$ (assuming the size is
fixed). Note that if we had tried to do better than
equation~\ref{eqn-D-I} and had replaced $\di[A]$ by
$i\!\not\!\partial$, then the eigenvalue spectrum of $V$ would tend to
that of the free fermion spectrum as the basis becomes complete.  This
is of course no reason to panic; this limit is clearly unphysical and
in any case all our simplifying assumptions will have certainly broken
down by then. However it explains why in our calculations, when we
vary the density, we find that at high densities our eigenvalue
spectra appear to tend towards $\propto \delta(\lambda-1/\rho)$.

To obtain some idea of the errors induced by this simplification, we
shall in at least one case below, use a more realistic version of the
Dirac operator.

\subsection{Zero-mode wave functions}

Since our calculations are in part numerical, we can only deal with a
finite number of instantons. That is to say, we will work in a finite
space-time volume. Of course, we are primarily interested in the $V
\to \infty$ limit and so we want the finite volume corrections to be
as small as possible. This can usually be achieved by working in a
periodic box, ${\mathbb T}^{4}$.  The wave-functions also then need to
be periodic. The calculation of the overlaps will usually be too
complicated to be performed analytically and will then need to be
performed numerically.  This calculation has to be performed to a very
high precision, otherwise our orthogonalisation procedure will
occasionally break down. With a large number of topological charges,
accurate numerical calculations of overlaps can become prohibitively
expensive. This means that in practice the wavefunctions that we can
use on a (hyper)torus are quite limited.

An alternative to working on ${\mathbb T}^{4}$ is to calculate
overlaps within ${\mathbb R}^{4}$ while locating the centres of the
instantons within a finite space-time volume $V$. As $V \to \infty$
one will obtain the same spectrum as one obtains by taking the volume
of ${\mathbb T}^{4}$ to infinity.  The drawback is that finite volume
corrections are likely to be much larger than for a torus. The obvious
advantage is that it is possible to do efficient calculations with a
much larger range of trial wavefunctions. For the sake of brevity we
shall refer to such calculations as being performed ``in ${\mathbb
R}^{4}$''.

We now list the various wave functions we shall be using in this
paper.

\subsubsection{Hard sphere}

The hard sphere wavefunction is defined by:

\begin{eqnarray}
\langle x|\psi_{Hj}^{\pm}\rangle & = & 1\qquad |x - x_{j}^{\pm}| \leq
\rho_{j}^{\pm}\nonumber\\
& = & 0\qquad \rm{otherwise},
\label{eqn-hard-sphere}
\end{eqnarray}
where $\rho_{j}^{\pm}$ is the measure of its size and $x_{j}^{\pm}$ is
its centre. This is the simplest wavefunction we use. In ${\mathbb
R}^{4}$ the overlap is simply given by the volume of intersection
between two solid balls in four dimensional space and this is easy to
calculate analytically.

If we are in ${\mathbb T}^{4}$ and if $\rho_j \leq l/2$, where $l$ is
the length of the torus, then the periodic copies of the hard sphere
do not overlap with each other. This greatly simplifies the
calculation of the overlap matrix element $U_{ij}$; we just calculate
the overlap between the j'th object and the i'th object as well as
with all the nearest periodic copies of the i'th object, and each of
these calculations can be performed analytically as though in
${\mathbb R}^{4}$. If $\rho_j \leq l/4$ things become even simpler
since the periodic copies then make no contribution.

Thus the calculations with $\psi_{H}(x)$ can be performed efficiently
in both ${\mathbb T}^{4}$ and ${\mathbb R}^{4}$. Hence it is the
wavefunction that was used in the calculations of \cite{ndmt} and in
\cite{usmt2}. It has some obvious drawbacks. In even a moderately
dilute gas there will be some instantons which have no overlap with
any anti-instantons and these will produce `accidental' exact
zero-modes. Moreover, one can easily show that in a dilute gas there
is a trivial $1/\lambda^{0.6}$ divergence
in the eigenvalue spectrum as $\lambda \to 0$,
that arises from the two-body eigenvalue splitting that results from
isolated (anti)instanton pairs whose overlap tends to 0.  Since a
divergence as $\lambda \to 0$ is one of the main issues in this paper,
we will need to perform careful checks to show that it does not have 
this trivial origin.

\subsubsection{Gaussian}

Gaussian wavefunctions in ${\mathbb R}^{4}$ have tails that stretch to
infinity and their overlaps can be calculated analytically. The
Gaussian wavefunction is given by:

\begin{eqnarray}
\langle x|\psi_{Gj}^{\pm}\rangle & = &
\frac{1}{\sqrt{2\pi}\sigma_{j}^{\pm}}\exp\left(\frac{(x-x_{j}^
{\pm})^{2}}{2\sigma_{j}^{\pm2}}\right).
\label{eqn-gaussian-r4}
\end{eqnarray}
where $x_{j}^{\pm}$ is the centre and $\sigma_{j}^{\pm}$ is a measure
of its size.

If one wishes to work in ${\mathbb T}^{4}$ then this wave-function
must be made periodic in some way. We shall choose to do so as
follows. First consider the following Gaussian function in ${\mathbb
R}^{4}$:

\begin{eqnarray}
G(x;x_{j}^{\pm},\sigma_{j}^{\pm},n) & = &
\frac{1}{\sqrt{2\pi}\sigma_{j}^{\pm}}\exp
\left(\frac{(x-x_{j}^{\pm}-n.l)^{2}}{2\sigma_{j}^{\pm2}}\right)\ ,
\end{eqnarray}
where $x_{j}^{\pm}$ lies in the volume $V=l^4$ and $n \in {\mathbb
Z}^{4}$. Then the following provides a periodic wave-function:

\begin{eqnarray}
\langle x|\psi_{{\widetilde{G}}j}^{\pm}\rangle & = & N\sum_{n \in
{\mathbb Z}^{4}}G(x;x_{j}^{\pm},\sigma_{j}^{\pm},n),
\label{eqn-gaussian-t4}
\end{eqnarray}
where $N$ is a suitable normalization constant. This is clearly a
smooth function on ${\mathbb T}^{4}$, since we have simply added all
its periodic copies to our original Gaussian.  The overlap of two such
wave-functions over ${\mathbb T}^{4}$ is simply the overlap over
${\mathbb R}^{4}$ divided by the number of ${\mathbb T}^{4}$
subvolumes contained in ${\mathbb R}^{4}$. Each such overlap is a sum
of terms each of which consists of the overlap between two simple
Gaussians defined on ${\mathbb R}^{4}$. This can be calculated
analytically. Hence our reason for choosing this particular periodic
extension of the Gaussian wavefunction.

Of course the above procedure involves infinite sums.  In practice,
however, a Gaussian falls off sufficiently fast with distance that we
can often ignore all but the nearest periodic copies, while maintaining 
a very good precision. When this is so the Gaussian calculation on
${\mathbb T}^{4}$ is only about $3^4$ times slower than the hard
sphere calculation. Thus while our largest scale calculations on the
torus will be with $\psi_H$, we shall be able to perform realistic
comparisons with $\psi_{\widetilde{G}}$ when required.

\subsubsection{Classical}

The classical zero-mode wave-function in ${\mathbb R}^{4}$ is defined
by:

\begin{eqnarray}
\langle x|\psi_{Cj}^{\pm}\rangle & = &
\frac{\sqrt{2}}{\pi}\frac{\omega_{j}^{\pm}}{(\omega_{j}^{\pm2} + 
(x - x_{j}^{\pm})^{2})^{\frac{3}{2}}} ,
\label{eqn-classical}
\end{eqnarray}
where $\omega_{j}^{\pm}$ is a size parameter and $x_{j}^{\pm}$ denotes
the centre of the object. In the semiclassical limit it would be the
correct wavefunction to use, but there is no reason to think that it
plays a special role in the full quantum vacuum. It has a slow
fall-off with distance and for this reason we do not use it on
${\mathbb T}^{4}$.  (The natural way to proceed on ${\mathbb T}^{4}$
would be to determine the approximate functional form of the zero-mode
numerically and to work with that.)

The overlap calculation is less trivial in this case than for the
hard sphere and Gaussian wavefunctions. Using the standard method
of Feynman parameters yields the following integral solution:

\begin{equation}
\langle\psi_{Ci} | \psi_{Cj}\rangle =
\frac{8}{\pi}\frac{\omega_{i}}{\omega_{j}}\int_{0}^{1}dq
\frac{\sqrt{q(1-q)}}{1
+ q(\frac{s^{2}}{\omega_{j}^{2}} + \frac{\omega_{i}^{2}}
{\omega_{j}^{2}}) - q^{2}\frac{s^{2}}{\omega_{j}^{2}}} .
\end{equation}
where $s^2 = (x_{i}- x_{j})^2$.The lack of a closed form expression 
for this overlaps makes the calculation far slower than when we use 
the hard sphere (or even the Gaussian) wavefunction. One can indeed 
evaluate this integral in terms of the hypergeometric function of 
two variables, but the quickest way of evaluating this function is 
to calculate the above one-dimensional integral numerically.

In the case of this wave-function we shall also perform some
calculations that approximate the covariant derivative by the partial
derivative, i.e.  $V_{kl} \simeq
\langle\psi_{k}^{+}|-i\!\not\!\partial|\psi_{l}^{-}\rangle$, rather
than by the identity, as in equation~\ref{eqn-D-I}.  There are several
parameterizations of this matrix element for classical zero mode
wavefunctions \cite{Shuryak,ShuVerb}, and we shall use the
parameterization:

\begin{eqnarray}
\langle\psi_{k}^{+}|-i\!\not\!\partial|\psi_{l}^{-}\rangle & \approx &
\frac{16R}{\rho_{k}^{+}\rho_{l}^{-}(2 +
R^{2}/\rho_{k}^{+}\rho_{l}^{-})^{2}}
\label{linadd}
\end{eqnarray}
where $R = |x_{k}^{+} - x_{l}^{-}|$ \cite{Shuryak}. We shall label
these calculations as ``type II classical''. This will provide a
non-trivial test of the robustness of our results against the extreme
approximation we have made for the covariant derivative.

\section{Quenched QCD (${\bf n_f}$=0)}
\label{section-results}

\subsection{Robustness}
The most striking result of the early work \cite{ndmt,ndmt2} with
(essentially) the above model was that the mixing between the would-be
instanton zero-modes produced a power-like divergence in the Dirac
spectral density as $\lambda \to 0$. The degree of divergence that was
found, $d \simeq 0.5$ in equation~\ref{eqn-power-law}, fitted in well
with the simple theoretical argument given in Section 2. The
calculations were, however, only performed with the hard-sphere
wave-function, $\psi_H$, and for a limited range of volumes and
packing fractions. Clearly one needs to test the robustness of this
divergence against varying the details of the calculation and this is
what we aim to do in this section.

The first question we address is: how large are finite volume
corrections when we use the hard-sphere wavefunction on ${\mathbb
T}^{4}$~?  This is the wavefunction, see
equation~\ref{eqn-hard-sphere}, that was used in \cite{ndmt}.  Before
presenting our results we need to choose some units in which to
express the volume. We choose our units so that the instanton size,
$\rho$, is 0.2 in those units.  That is to say, our unit of length is
$5 \times \rho$.  These are the units we shall use throughout the
present paper. We shall show the results of the finite volume study
for a packing fraction of unity, $f=1$. This is a value intermediate
between the extremes of very high and very low density where our
approach is certain to break down (as discussed earlier). It is a
natural value to consider in the pure gauge theory, which has only one
scale. Note that for $f=1$ a volume $V=1$ will contain
$n_{\overline{I}} + n_{I} = 126$ topological objects. In our
calculations we shall, unless explicitly stated otherwise, use
$n_{\overline{I}} = n_{I}$ so that $Q=0$, in the expectation (as
discussed earlier) that this will minimise finite-$V$ corrections.

\begin{figure}[p]
\begin{center}
\leavevmode
\epsfxsize=100mm
\epsfbox{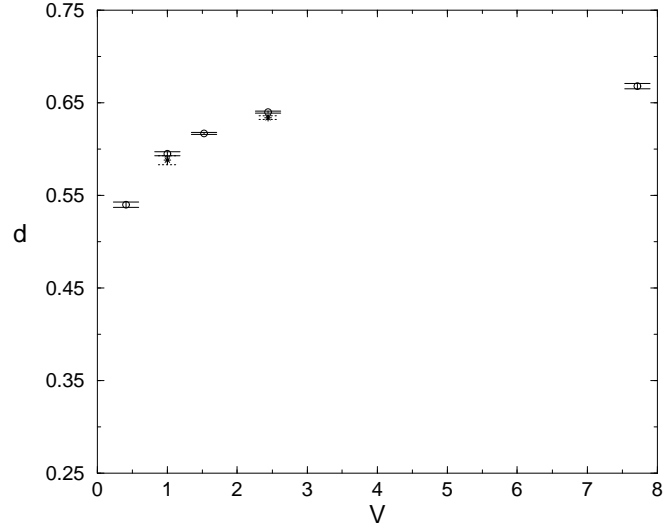}
\caption{Exponent of power divergence, $d$, versus space-time volume
$V$. Hard sphere ($\circ$) and Gaussian ($\star$) wavefunctions, 
with density $f=1$.}
\label{fig:hg_dvsV}
\end{center}
\end{figure}

\begin{figure}[p]
\begin{center}
\leavevmode
\epsfxsize=100mm
\epsfbox{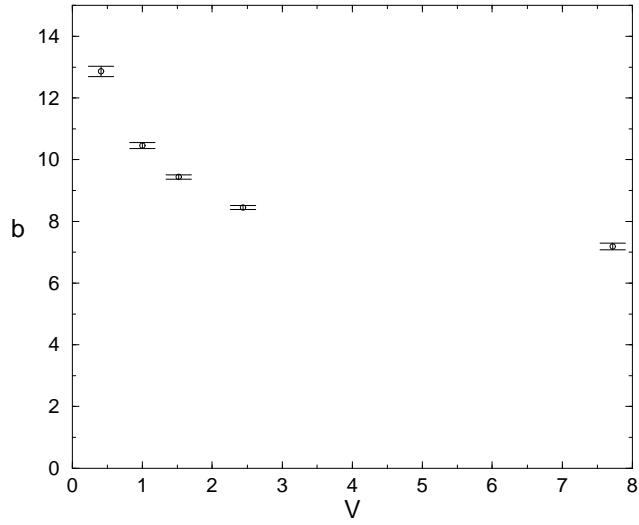}
\caption{Volume dependence of the coefficient $b$ of the 
$1/\lambda^d$ piece in the spectral density. For hard spheres 
as in figure~\ref{fig:hg_dvsV}.}
\label{fig:hg_dvsV_coeff}
\end{center}
\end{figure}

So, we have calculated the Dirac spectra for various values of $V$ at
$f=1$ and for $Q=0$. We have fitted the small $\lambda$ part of the
spectrum to the form in equation~\ref{eqn-power-law}. In
figure~\ref{fig:hg_dvsV} we show how the degree of divergence, $d$,
varies with the volume $V$. In figure ~\ref{fig:hg_dvsV_coeff} we show
the corresponding variation of the coefficient of the divergent term
as a function of the volume (i.e. $b$ in
equation~\ref{eqn-power-law}).  We see that both quantities appear to
have finite non-zero limits as $V \to \infty$, and that at $V=1$ the
corrections are modest; so such a volume can be used to extract
qualitative features of the spectrum.  Although we do not show the
full spectra we remark that finite-$V$ corrections are small across
the whole range of $\lambda$ and not just for $\lambda \sim 0$.  We
have also performed some calculations on different space-time volumes
at other values of $f$ and the conclusions, of these less systematic
studies are in agreement with what we have found at $f=1$. That is to
say, finite volume corrections are small and, for our qualitative
purposes, one can safely work on a space-time volume $V=1$ in our
units.

\begin{figure}[p]
\begin{center}
\leavevmode
\epsfxsize=100mm
\epsfbox{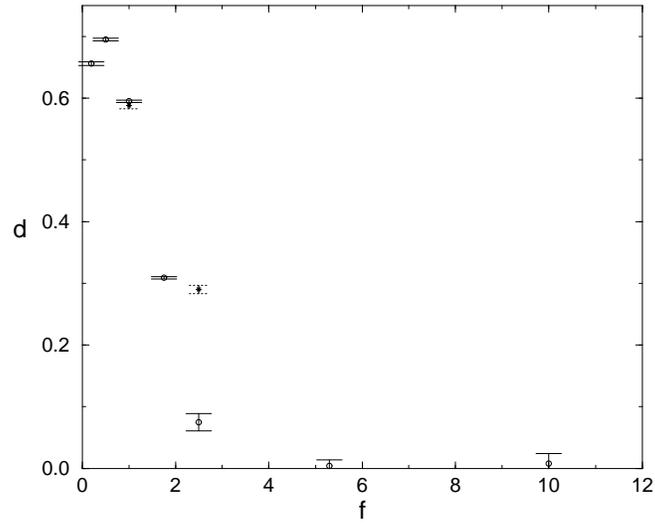}
\caption{Exponent of power divergence, $d$, versus packing fraction
$f$. Hard sphere ($\circ$) and Gaussian ($\star$) wavefunctions, 
with volume $V=1$.}
\label{fig:hg_dvsf}
\end{center}
\end{figure}

The second question we ask is: how does the degree of divergence $d$
depend on the packing fraction, $f$, as defined in
equation~\ref{eqn-pack-frac}? We perform our calculations on ${\mathbb
T}^{4}$ using hard sphere wavefunctions with $Q=0$ and on a volume
$V=1$. The results are shown in figure ~\ref{fig:hg_dvsf}. We observe
that there is a rapid decrease of $d$ with increasing instanton
density precisely in the range of densities that is of physical
interest. Thus, while we can claim that the existence of a divergence
is a characteristic feature of the instanton induced Dirac spectrum,
the precise degree of divergence is not something that we can claim to
predict with our model.  We note that at very high densities the
divergence effectively disappears. However, since our model becomes
unreliable in that limit we should be cautious about inferring that
this is the case for the true spectrum. We also note that at very low
densities the value of $d$ will be determined by the detailed nature
of the tail of the wave-function being used.

The calculations described above are specified in more detail in
table~\ref{tab:basicdat}. We show there the $\chi^2/N_{DF}$ (per
degree of freedom) of the best fit. We have typically fitted the
eigenvalue range $\lambda \in [0,0.3]$ where the best fits have
$\chi^2/N_{DF}$ between 1 and 2.  (Note that with our choice of units,
the maximum value of $\lambda$ is about 5.0). In practice one can
easily reduce the value of $\chi^2/N_{DF}$ to less than unity by
reducing the fitted range to about $\lambda \in [0,0.1]$. The value of
$d$ is almost completely insensitive to which of these two ranges is
chosen. In either case, the range of $\lambda$ is substantial enough
that the spectrum therein is determined with a high accuracy so that
the fact that equation~\ref{eqn-power-law} fits well provides
convincing evidence that we do indeed have a power-like divergence as
$\lambda \to 0$.

To emphasise the last point further we have also fitted our $\lambda
\to 0$ spectra with a logarthmic divergence:

\begin{equation}
\overline{\nu}(\lambda) = a + b \ln(\lambda) .
\label{eqn-log-law}
\end{equation}
Typical best $\chi^2/N_{DF}$ values are shown in
table~\ref{tab:basicdat}. We see that such fits are quite unacceptable
except at the higher densities where the divergence almost disappears,
$d\to 0$. In that case the trivial expansion $\lambda^{-d} \simeq 1-d
\ln(\lambda)$ explains the inevitable compatibility of both kinds of
fits. We remark that a logarithmic divergence is of particular
interest because of its appearance in quenched chiral perturbation
theory \cite{cpt-log} and it has been recently discussed in the
context of Random Matrix Theory \cite{dvgce-other}. For a brief
discussion of the possible connection between our power-like
divergence and that in quenched chiral perturbation theory
\cite{cpt-power} we refer the reader to the concluding section of
\cite{usmt2} and of this paper.

We now turn to two more complex questions: first, how robust are the
qualitative features we have described above against changes in the
functional form of the would-be zero-mode wavefunction, and second,
are these features in some way special to our particular
simplification of the Dirac operator?

\begin{figure}[p]
\begin{center}
\leavevmode
\epsfxsize=100mm
\epsfbox{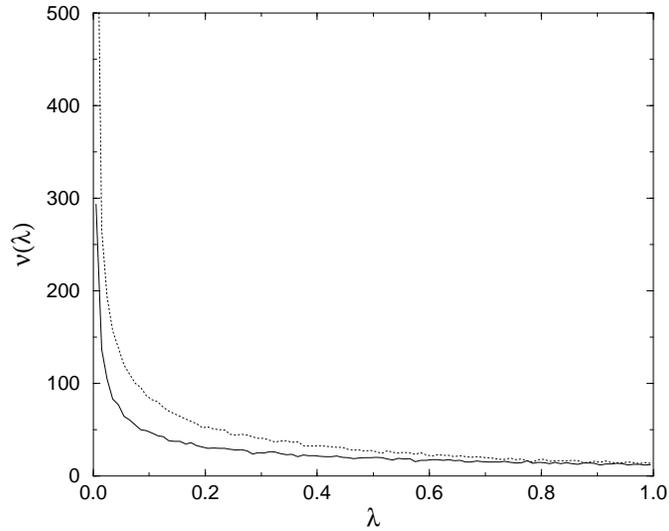}
\caption{Spectral density using classical zero mode wavefunctions. 
Our usual approximation to \di[A] (solid) compared to that using
the linear addition ansatz (dotted).}
\label{fig:cs}
\end{center}
\end{figure}

First, the Dirac operator. One thing we can do (and the only thing we
will do) is to see what happens if one goes one step `better' than we
have done so far and replaces the covariant derivative by the partial
derivative as in equation~\ref{eqn-D-to-d}. We use the `classical'
form for the zero-mode, as in equation~\ref{eqn-classical}. We then
generate two ensembles (in ${\mathbb R}^{4}$) with the same values of
$f$ and $V$. (They happen to be $f \simeq 1$ and $V \simeq 1$ but we
have not yet discussed how we determine $f$ and $V$ for anything other
than the hard-sphere wavefunction.)  The first ensemble uses our usual
simplification for the covariant derivative, as in
equation~\ref{eqn-D-I}. The second ensemble calculates the matrix
element of the derivative using the approximate form given in
equation~\ref{linadd}. We compare the resulting Dirac spectra in
figure~\ref{fig:cs} and table~\ref{tab:basicdat} (with the latter
ensemble under the heading ``Class. II''). We see that while there are
some quantitative differences, as there must be, the qualitative
features -- chiral symmetry breaking and a divergence as $\lambda \to
0$ -- are certainly both there.  Indeed the values of the degree of
divergence as listed in table~\ref{tab:basicdat} are almost
identical. This provides some non-trivial evidence that the
qualitative features we are emphasising, do not in fact arise from our
particular simplification of the covariant derivative.

\begin{figure}[p]
\begin{center}
\leavevmode
\epsfxsize=100mm
\epsfbox{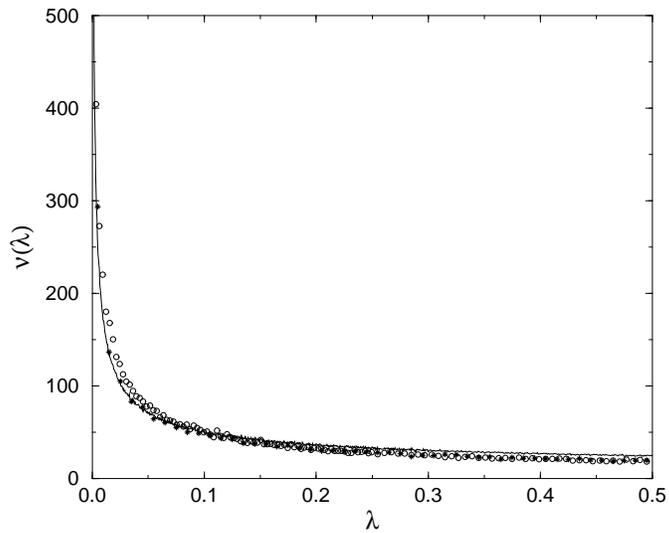}
\caption{The spectral densities obtained at $f=1,\ V=1$ from hard
sphere (solid), Gaussian ($\circ$) and classical zero mode ($\star$)
wavefunctions.}
\label{fig:hgc}
\end{center}
\end{figure}

How robust are our results against changing the wavefunction?  A
preliminary answer to this question is provided in
figure~\ref{fig:hgc}. There we plot the spectral densities obtained
with hard-sphere, Gaussian and classical zero-modes respectively, in a
volume $V=1$ and with a packing fraction, of $f=1$ for the
hard-sphere. We observe that they are indeed nearly identical.
 
\begin{figure}[p]
\begin{center}
\leavevmode
\epsfxsize=100mm
\epsfbox{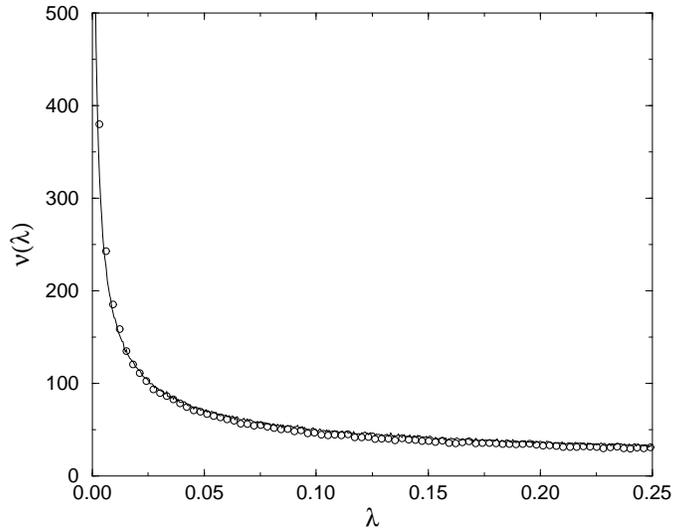}
\caption{Comparing spectral densities on ${\mathbb R}^{4}$ ($\circ$)
and on ${\mathbb T}^{4}$ (solid), with hard sphere wavefunctions 
for $f=1,\ V=1$.}
\label{fig:r4_t4}
\end{center}
\end{figure}

Of course, this comparison skates over some points that need
elaboration. Firstly, while the hard-sphere and Gaussian wavefunction
calculations are in ${\mathbb T}^{4}$, the calculation with the
classical wavefunction is in ${\mathbb R}^{4}$. Are the spectra
obtained in ${\mathbb T}^{4}$ and in ${\mathbb R}^{4}$, for a given
wavefunction choice, similar enough for such a comparison to be
useful? Of course they must become identical as $V \to \infty$; but at
$V=1$?  To answer this question we show in figure~\ref{fig:r4_t4} the
spectral densities that one obtains with $f=1$, $V=1$ hard-sphere
ensembles in ${\mathbb T}^{4}$ and ${\mathbb R}^{4}$ respectively. We
observe that they are very similar. We thus conclude, from this and
some similar comparisons, that a volume of $V=1$ is already large
enough to allow comparisons on these two, somewhat different
space-time manifolds.

A second point concerns the `packing fraction' $f$ of wavefunctions
other than the hard sphere. We have seen that the hard-sphere Dirac
spectrum varies rapidly with $f$, and so this is likely to be the case
for the other wavefunctions.  So ideally one would want to plot say
$d$ against $f$ and see if the results were qualitatively the same for
all wavefunctions.  Unfortunately there cannot be a definiton of $f$
that is really equivalent for all the wavefunctions. So the best we
can do is to see if one can actually find values of the size
parameters $\sigma$ and $\omega$ such that the divergences obtained
with Gaussian and classical wavefunctions are the same as the one
obtains with hard spheres, for the chosen values of $f$ and $V$. If
that is the case -- and if so it would represent a non-trivial result
-- then it will be interesting to see if the various ensembles possess
some other properties that indicate that they are of a comparable
density.

We therefore proceed as follows. We consider some hard sphere
configurations, in a fixed volume $V$, fixed packing fraction $f$
consisting of objects of a single size $\rho$. To be concrete let us
choose $V=1$ (the unit four torus), $\rho=0.2$ and $n_{\overline{I}} +
n_{I} = 63 + 63$ which gives $f=1$. We extract the spectral density
from these configurations and fit it using the power law form given in
equation~\ref{eqn-power-law}.  We then generate configurations using a
Gaussian wavefunction of size $\sigma$ and with the same number of
topological objects as before. The latter ensures that when $\sigma$
is equivalent to $\rho=0.2$, the volume will be $V=1$. We now extract
the spectral density, do a power law fit for $\lambda \to 0$ and
extract a value of the divergence exponent $d$. We repeat this for
various values of $\sigma$. If we find a value for which $d$ is the
same as for the hard-sphere case, then we say that this is the
equivalent value. (Note that when we vary $\sigma$ we not only vary
$f$ but we also vary $V$. However we have seen for the hardsphere
wavefunction that the variation of $d$ on $V$ is weak in comparison to
its variation on $f$.)  We repeat the above using the classical
zero-mode.  If we follow this procedure we find that $\sigma=0.074$
and $\omega=0.02$ are equivalent to $\rho=0.2$. These values were in
fact the ones used to generate the spectra shown in
figure~\ref{fig:hgc}. Some properties of the various spectra are
listed in table~\ref{tab:basicdat}. There we label the various
ensembles by the `equivalent' $f$ and $V$ as determined above.

The fact that we tuned the size parameters $\sigma$ and $\omega$ so
that they would give the same divergence as the hard sphere spectrum
with $\rho=0.2$ does not mean that the comparison in
figure~\ref{fig:hgc} is in any way trivial.  The fact that one can
successfully perform such a tuning already shows that power-law
divergences naturally arise with all these quite different
wave-functions. We also see in figure~\ref{fig:hgc} that when we
perform this tuning, the whole spectrum, and not just the divergent
piece, is quite robust against changes in the wavefunction.

Using the above way of setting a common scale for the various
wavefunctions, we can extend our earlier check of the $V$ dependence
of the divergent $\lambda \to 0$ piece of the spectrum. We do this for
the Gaussian wavefunction in figure~\ref{fig:hg_dvsV}. We see that the
finite-$V$ corrections are very much as for the hard sphere. In figure
~\ref{fig:hg_dvsf} we show how $d$ varies with the packing fraction
$f$ for the Gaussian. We see that, just as for the hard sphere, the
divergence becomes weaker at higher densities. However the agreement
is only qualitative. This means that if we had performed our `tuning'
to the hard-sphere spectrum at a different value of $f$ than $f=1$,
then we would have obtained a somewhat different equivalent value of
$\sigma$. This is inevitable since different wavefunctions are
certainly not equivalent in all respects.

It is interesting to check, in some quite independent way, whether the
above `equivalent' sizes do really correspond to packing fractions
that are very roughly comparable.  Consider, for example, the total
overlap of an instanton would-be zero-mode with all the anti-instanton
would-be zero modes in the configuration. This will clearly increase
linearly with the packing fraction (since the charges are
positioned at random) and will be independent of $V$ once $V$ is large
enough. So its value clearly provides a measure of the density that
allows a crude comparison between different wavefunctions. We have
calculated this integrated overlap for the hard-sphere, Gaussian and
classical ensembles used in figure~\ref{fig:hgc} and find values
0.496, 0.297 and 0.495 respectively. These are indeed similar enough
to be reassuring.

\subsection{Realistic instanton ensembles}

One might expect that a simple way to learn what actually 
occurs in quenched QCD, is to replace the artificial ensembles 
that we have been using, in most of which the instantons have 
a fixed size and number and are positioned at random, with the 
instanton ensembles extracted in actual lattice simulations
\cite{DSMT,qlat-other}. Unfortunately such an approach
possesses ambiguities both in principle and in practice. An 
example of the former is the question of how to separate 
the effect of topological from  non-topological fluctuations;
after all, the detailed featrures of the instanton ensemble
will be in part due to such non-topological fluctuations.
A practical difficulty one encounters\cite{usmt2} is the fact
that the density of the lattice instanton ensembles turns
out to be very sensitive to the smoothing of 
the gauge fields, which is necessary to remove the high-frequency
fluctuations that would otherwise obscure the (lattice) topological 
charge density. Depending on how much smoothing one performs, one 
can obtain \cite{usmt2} spectral densities with or without a 
small-$\lambda$ peak.

That is not to say that calculations with realistic instanton
ensembles don't have interesting lessons to teach us. For example one
learns \cite{usmt2} that within an apparently dense instanton `gas',
the narrower instantons can contribute a small-$\lambda$ peak to the
Dirac spectral density, just as though they were part of a much less
dense gas. The reason for this phenomenon is simple \cite{usmt2}. The
high packing fraction of the lattice instanton `gas' is dominated by
the very large instantons (since $V_I \propto \rho^4$ in
equation~\ref{eqn-pack-frac}) but such instantons will have a small
part of their wavefunction in the small region of space-time
occupied by the wavefunction of a small instanton. Thus the matrix
elements of \di\ between large and small instantons can be neglected
to a first approximation, and the latter effectively interact as
though they belonged to a much less dense gas of, on the average,
smaller instantons. This observation is of particular interest
because it (potentially) provides a justification for the 
use of relatively dilute gases of small instantons in various 
successful phenomenological studies\cite{Shuryak}.

In \cite{usmt2} the would-be zero mode wavefunctions used were hard
spheres. However we have seen in this paper that while we can set up
an approximate equivalence between different wavefunctions by matching
spectral densities or, more directly, the overlaps themselves, the
actual scales characterising the matched wavefunctions are quite
different. Since we have not determined which scale corresponds to a
particular instanton size as determined in a lattice calculation, this
again leaves a great deal of ambiguity in determining the
corresponding packing fraction, $f$, since doing so would require 
determining the matched hard-sphere radius to be used in
equation~\ref{eqn-pack-frac}.  This is an ambiguity 
of which the reader of \cite{usmt2} should be aware.

Indeed when we refer in this paper to the size of an instanton, for
example $\rho=0.2$, the reader should be aware that what we mean is
that the instanton has such a size that its would-be zero mode can be
approximated by a hard-sphere of radius $\rho=0.2$. (That the real
would-be zero-mode can be approximated by a hard sphere is suggested
by the fact that we can match the spectral density from hard spheres
with that from Gaussians or classical zero modes.)  Its actual size,
in terms of the spread of the topological charge density, may well be
somewhat different, and trying to determine that (for the fully
fluctuating vacuum of interest) lies well beyond the scope of 
the present work.

\subsection{Dynamics}

We have provided evidence that the power-like divergence induced in
the Dirac spectrum by instantons is not an artefact of the
approximations used in our model but is a real effect.  We can
therefore begin to ask some questions about its dynamical
origins. There are three issues we shall address. The first is whether
there is any evidence for the simple picture \cite{ndmt} for the
origin of this divergence that we presented in Section
~\ref{section-chiral}. The second question is whether the divergence
has an origin in the simplest two-body interactions between instantons
and anti-instantons. Finally we ask whether the eigenvectors of the
Dirac operator have any striking features as $\lambda \to 0$.

\subsubsection{Eigenvalue replacement hypothesis}

In Section~\ref{section-chiral} we described a simple scenario in
which a divergent spectrum will naturally arise. In that argument one
assumes that the only finite-$V$ correction in the (normalised) Dirac
spectrum consists of a cut-off for $|\lambda| \leq c/V$ (for some
constant $c$). If we increase the volume from $V$ to $V + \delta V$
then the (fractionally) greater number of modes in the interval
$|\lambda| \in [c/(V + \delta V) , c/V ]$ must match the fractional
decrease in the number of exact zero-modes.  This assumption that the
fractional decrease in the exact zero modes is compensated for by an
increase amongst the very smallest non-zero modes, we earlier referred
to as the `eigenvalue replacement hyopothesis'. As we saw, it leads
to a divergent spectrum, as $\lambda \to 0$, with a power exponent
$d=0.5$. The fact that in practice we find that the power depends on
the packing fraction tells us that this argument is approximate at
best. It is interesting nonetheless to see if it does capture some of
the essential features of what is happening.

A simple first step (which is as far as we shall try to go in this
paper) is to determine whether our eigenvalue replacement hyopothesis
is qualitatively correct. Since this concerns the nature of the
finite-$V$ corrections, we shall need to generate configurations with
no constraint on the value of $Q$. Our strategy will be
twofold. Firstly we generate configurations at a fixed packing
fraction but for two volumes, and we check explicitly whether the
finite volume corrections to the spectrum are really concentrated at
small values of $\lambda$. Secondly we shall perform calculations at
fixed $V$ and compare, for example, configurations with $Q=0$ and
$Q=2$ so as to see whether the extra 2 non-zero modes in the former
are indeed concentrated amongst the lowest modes of the latter.

\begin{figure}[p]
\begin{center}
\leavevmode
\epsfxsize=100mm
\epsfbox{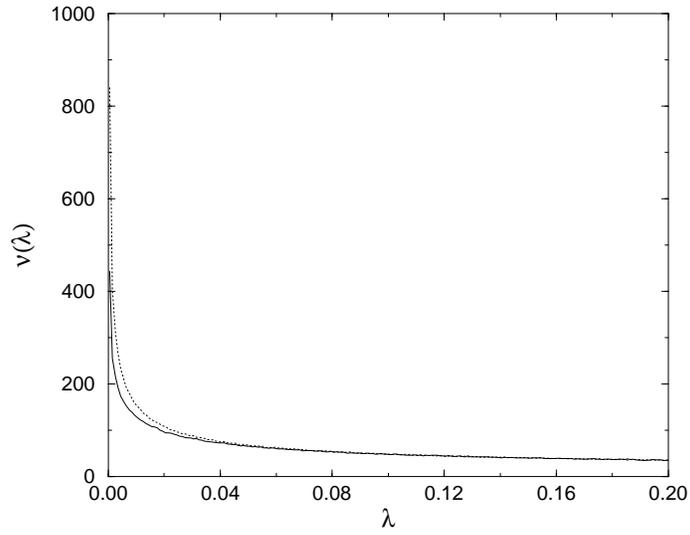}
\caption{Spectral densities with variable $Q$ on
two volumes:  $V=1$ (solid) and $V\approx 2.44$ (dashed).
With hard spheres and $f=1$.}
\label{fig:varyq}
\end{center}
\end{figure}

\begin{figure}[p]
\begin{center}
\leavevmode
\epsfxsize=100mm
\epsfbox{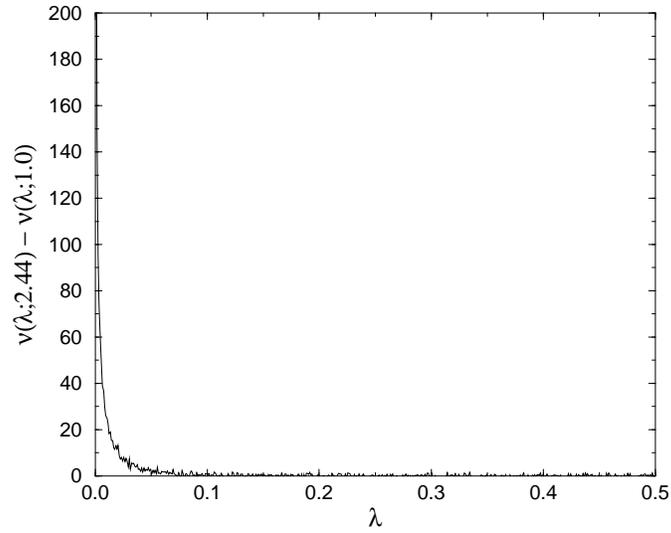}
\caption{The difference of the two spectral densities in
figure~\ref{fig:varyq}.}
\label{fig:varyq_d}
\end{center}
\end{figure}

For our first calculation, the one with two different volumes, we draw
the number of instantons and anti-instantons in each configuration
from a normal distribution:

\begin{equation}
N_{I,A} \sim N(\bar{N}/2, \bar{N}/4)
\label{eq:varyia}
\end{equation}
where $\bar{N} \propto V$ is the mean number of topological charges in 
the volume $V$ (for instance we have $\bar{N}=126$ for $V=1,\ f=1$). We 
see therefore that the total number of objects $N = N_{I} + N_{A} \sim
N(\bar{N}, \bar{N}/2)$ and that the winding number distribution follows $Q
= N_{I} - N_{A} \sim N(0, \bar{N}/2)$. (The actual numerical factors
arise as this is nothing other than the central limit theorem applied
to the binomial distribution $N_{I,A} \sim B(\bar{N},1/2)$ for large
$\bar{N}$.) The resulting Dirac spectra for such a gas in two different
volumes is shown in figure~\ref{fig:varyq} and is listed in
table~\ref{tab:varyq}. The results clearly show a far greater
difference at small eigenvalues with the curves converging for
$\lambda \geq 0.04$. (Note that the spectrum extends out to
$\lambda_{max} \sim 5$.) This can be seen more clearly if we simply 
focus on the difference between the two spectral densities as in
figure~\ref{fig:varyq_d}. This provides qualitative support for what
we have called the eigenvalue replacement hypothesis.

As an aside, we note from table~\ref{tab:varyq} that the value of $d$
for ensembles where $Q$ varies shows much stronger finite-$V$
corrections than for the $Q=0$ ensembles listed in
table~\ref{tab:basicdat}. (As $V \to \infty$ both ensembles should
give the same value of $d$.) This shows that our earlier strategy of
using $Q=0$ ensembles so as to minimise finite volume corrections was
appropriate.

One way to go beyond the above qualitative analysis is as follows.
Consider configurations with some fixed total number of charges
i.e. in a fixed volume.  We denote the un-normalised density of {\it
non-zero} eigenvalues obtained from configurations with topological
charge $\pm Q$ by $\rho(\lambda, |Q|)$. Note that we explicitly
exclude the exact zero-modes from this density. Its integral over the
interval $[\lambda_1,\lambda_2]$ is simply the average number of
eigenvalues in that interval for such configurations.  Let us now
define a second density, $\rho_{min}(\lambda, |Q|)$, that counts only
the smallest positive eigenvalue in each configuration. So its
integral over $\lambda \in [0,\infty]$ is just 1. Suppose we now
simultaneously consider configurations with charge $|Q|$ and
$|Q|+2$. The latter will have 2 extra zero-modes, i.e. the former will
have one extra positive mode (and of course a corresponding negative
mode as well). In this context the eigenvalue replacement hypothesis
would say that the two extra zero-modes in the configurations with
charge $|Q|+2$ replace the lowest positive (and largest negative)
eigenvalues in the configurations with charge $|Q|$, i.e.

\begin{equation}
\rho_{min}(\lambda,|Q|) = \rho(\lambda,|Q|) - \rho(\lambda,|Q|+2).
\label{eqn-rep-hyp}
\end{equation}
We shall test the following integrated form of this relation.  The
smallest eigenvalue is obviously localised at small $\lambda$.  Define
$\lambda_0$ by

\begin{equation}
\int\limits_0^{\lambda_0} \rho_{min}(\lambda,|Q|) = 0.98.
\label{eqn-rep-hyp2}
\end{equation}
Thus the interval $[0,\lambda_0]$ is essentially where the lowest
positive eigenvalue is localised. (We have used 0.98 rather than 1 in
equation~\ref{eqn-rep-hyp2} because formally at least there will be a
non-zero, even if infinitesimal, probability for the smallest
eigenvalue to be large.) We can now calculate the quantity

\begin{equation}
\delta\rho(|Q|,|Q|+2) = 
\int\limits_0^{\lambda_0}
\{\rho(\lambda,|Q|) - \rho(\lambda,|Q|+2)\}.
\label{eqn-rep-hyp3}
\end{equation}
If the eigenvalue replacement hypothesis held exactly then we would
expect this to be 0.98 for all $|Q|$. This test was performed in
\cite{ndmt2} using hard spheres and we quote the results of that
calculation in table~\ref{try_ndmt}.  (In \cite{ndmt,ndmt2} the
overlap calculation is cruder than ours, but that should not make a
significant difference.) We observe from the table that the eigenvalue
replacement hyothesis does indeed provide an approximately valid
description.  Typically about $80\%$ of the difference between the
$|Q|$ and $|Q|+2$ spectra resides in the region occupied by the very
lowest eigenvalue in the configurations with $|Q|$.  We also note that
as $|Q|$ increases this percentage decreases.  It is presumably
this feature that leads to the power of the divergence not being
precisely $0.5$ and is what enables it to vary with the packing 
fraction. Thus, while the
eigenvalue spectra do not satisfy the eigenvalue replacement
hypothesis exactly, they do so approximately and it presumably
provides a qualitative starting point for understanding the origins of
the divergence.

\subsubsection{Two body interactions}

In the case of hard-sphere wavefunctions, when the gas is sufficiently
dilute that the chance for an instanton to overlap with more than one
other charge may be neglected, it is easy to show analytically that
the eigenvalue spectrum diverges as $\lambda \to 0$ as
$1/\lambda^{0.6}$. This follows from the fact that if two hard spheres
overlap by a small distance $\delta$ then the overlap volume is
$\propto \delta^{5 \over 2}$.  This will then be the eigenvalue shift
from 0 due to the two body mixing. The probability measure is $\propto
d\delta$, which, because $\lambda \propto \delta^{5 \over 2}$, is just
$\propto 1/\lambda^{0.6}$. Thus here the origin of the divergence is
relatively trivial. The fact that our observed divergence appears for
all the quite different wavefunctions that we have used, already tells
us that its origin can hardly be so trivial. Nonetheless it does raise
the question whether it is primarily due to 2-body dipole-like mixing,
in which case it should be easy to understand, or to much more complex
multi-body mixings. This is the question we shall now address.

We address this issue in two ways. Firstly we generate a `background
spectral density' consisting of simple pairwise splittings of
eigenvalues. Secondly, we explicitly construct a gas of dipoles and
see if we still find a peak at small eigenvalues.

\begin{figure}[p]
\begin{center}
\leavevmode
\epsfxsize=100mm
\epsfbox{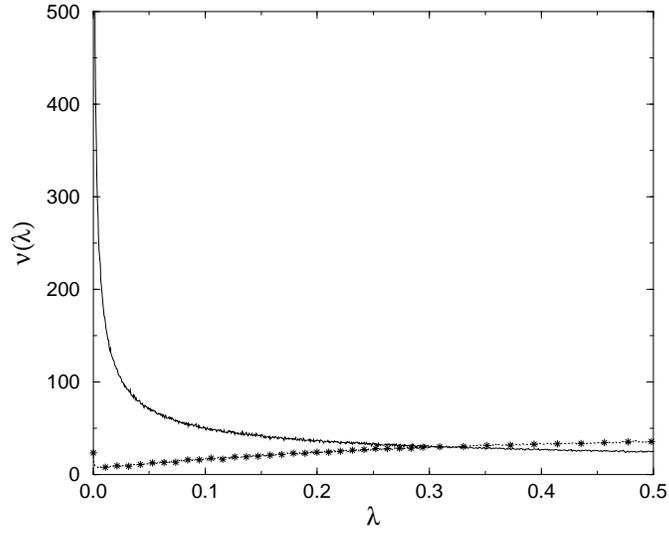}
\caption{Actual spectral density (solid) compared to
background curve ($\star$) from pairwise eigenvalue splitting
as defined in the text. Hard spheres with $V=1,\ f=1$.}
\label{fig:hbck}
\end{center}
\end{figure}

Figure~\ref{fig:hbck} shows the spectral density for the hard sphere
wavefunction coming from the actual eigenvalues of our Dirac
matrix. It also shows a `background curve' which originates from
pairwise interactions between the topological charges in the same
ensemble calculated as follows. Consider a configuration with $Q=0$. 
We go through the set of instantons and for each one find the largest
overlap with a neighbouring anti-instanton.  We then use this overlap
in a 2-body mixing matrix to obtain the eigenvalue shift from zero:

\begin{equation}
\lambda_{i} = \pm \max_{j} \{V_{ij}\}
\end{equation}
This then produces for us the `background spectral density' plotted in
figure~\ref{fig:hbck}. We remark that if we had $Q < 0$ then we would
again go through the $n_{I}$ instantons, associating each with two
eigenvalues, symmetrically split about zero, as above. The rest of the
eigenvalues are zero by the the Atiyah-Singer theorem. Analogously for
$Q > 0$.  The background curve thus generated only takes into account
pairwise splitting and furthermore ignores totally the effects of
other objects of the same charge in the vicinity. We see that the
$\lambda \to 0$ peak does not appear in the background curve. This
indicates that the $\lambda \to 0$ peak that we observe in the Dirac
spectrum is not due a simple 2-body interaction.

There are many ways to produce such a background curve, so this
example cannot be considered conclusive. For example one might try to
produce a background curve just as above except that any
anti-instanton that is used once in a 2-body mixing cannot be used
again; it is effectively removed. This produces a quite different
spectrum, which has a plethora of extra near-zero and exactly zero
eigenvalues. This is because, as we remove anti-instantons from the
configuration, we are soon left with a very dilute gas of topological
charges. This is clearly an artefact of the procedure, but serves to
illustrate the inherent ambiguity in trying to generate a background
curve.

\begin{figure}[p]
\begin{center}
\leavevmode
\epsfxsize=100mm
\epsfbox{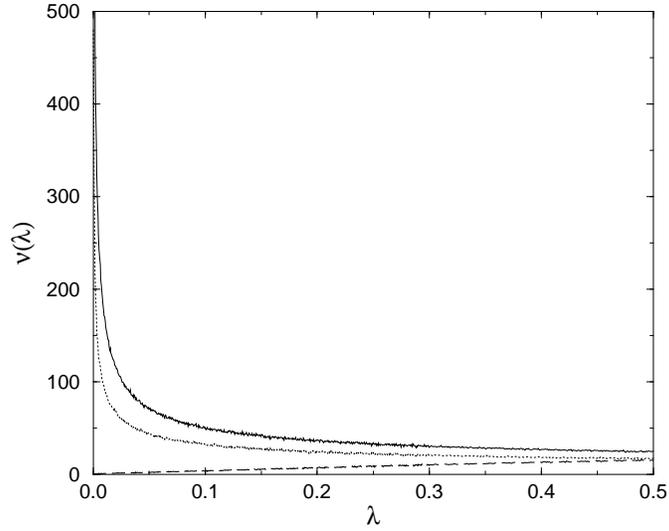}
\caption{Spectral densities: instanton gas (solid), dipole gas with
maximum separation $|x^{+} - x^{-}| \leq 2\rho$ (dotted);
background curve for dipole gas (long dashed). Using hard
spheres and $V=1,\ f=1$.}
\label{fig:dip1}
\end{center}
\end{figure}

To further explore this question we explicitly construct a $Q=0$
ensemble of topological charges from a gas of dipoles, pairing off
each object with one of the opposite chirality. We use hard sphere
wavefunctions of size $\rho$ and the distance between the opposite
charges within each dipole is chosen at random (with a four-volume
weighting) within some interval $|x^+ - x^-| \in [0, r_{max}]$ for
some $r_{max}$.  If $r_{max} < 2\rho$ then the dipoles will always
have a non-zero overlap and we would naively expect no eigenvalues
near zero if only 2-body interactions are important.  In
figure~\ref{fig:dip1} we show the spectrum for such a gas of dipoles
in the case where $r_{max} = 2\rho$. We compare it to the spectrum one
obtains with a $Q=0$ gas of (anti)instantons placed at random in the
volume. We also show the background spectrum that one obtains by
calculating the eigenvalues of the dipole gas, from the mixing within
each dipole taken in isolation. (Note that this may differ somewhat
from our previously defined background spectrum.) We observe that the
dipole gas spectrum is far from the background curve and not so
different to that of the randomly placed charges.  This shows that the
eigenfunctions corresponding to the small modes in the divergent peak,
are formed by a mixing that is very far from simple two body.

\begin{figure}[p]
\begin{center}
\leavevmode
\epsfxsize=100mm
\epsfbox{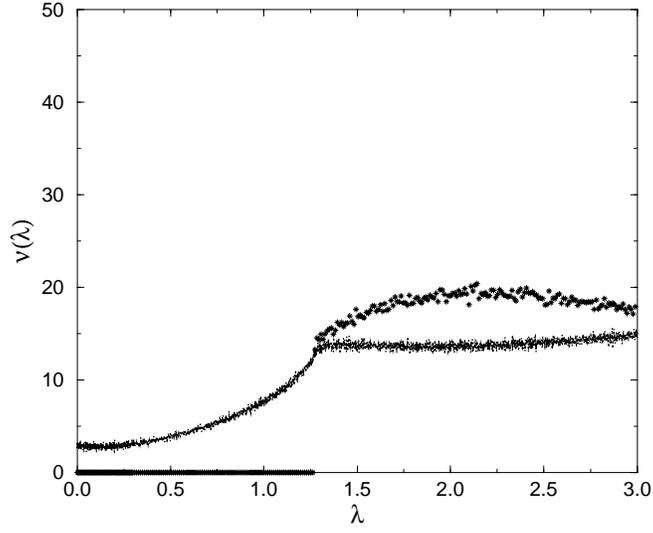}
\caption{Actual spectral density (dotted) compared to
background curve ($\star$) for a dipole gas with
maximum separation $|x^{+} - x^{-}| \leq 2\rho$. With $V=1,\ f=1$.}
\label{fig:dip2}
\end{center}
\end{figure}

An even more striking example is provided by choosing $r_{max} =
\rho$. Now the minimum overlap is very large and we would naively
expect the spectrum to have no small eigenvalues at all. In
figure~\ref{fig:dip2} we show the background curve obtained by
treating each dipole in isolation, and indeed it possesses a sharp cut
off as expected. Remarkably enough, however, the actual Dirac spectrum
we calulate from this gas of heavily overlapping dipoles does possess
a non-zero density of modes at $\lambda = 0$ and hence breaks chiral
symmetry spontaneously. This shows that there is a component
to the mixing that is far from simple two-body, and it is this
component that drives the spontaneous breaking of the chiral symmetry.

\subsubsection{The eigenfunctions for $\lambda \to 0$}

It would obviously be interesting to learn something about those
eigenfunctions of the Dirac operator that correspond to the small
eigenvalues that drive chiral symmetry breaking. We have seen above
that their origin lies in something more complex than simple two-body
mixing.  Since chiral symmetry breaking is associated with the
presence of a massless Goldstone boson, the pion, one might expect
that the eigenfunctions would become more extended as $\lambda \to 0$;
perhaps, even, that the extent of these eigenfunctions might
diverge. If this were so, then this could perhaps be linked to the
masslessness of the pion  using the standard decomposition of the 
component quark propagators in terms of the eigenvalues and 
eigenfunctions of the Dirac operator \cite{kogut98}.

To explore this question we need a measure of the dispersion of the 
eigenfunction $|e_{\lambda}\rangle$ corresponding to the eigenvalue 
$\lambda$ of our Dirac operator, i.e. \di$|e_{\lambda}\rangle =
\lambda|e_{\lambda}\rangle$.  We first define the function

\begin{equation}
R(x_{c}) = \langle e_{\lambda}|(x - x_{c})^{2}|e_{\lambda}\rangle ,
\end{equation}
that is clearly a measure of the dispersion around the point $x_c$. 
The extent space-time of the eigenfunction is provided by the dispersion
about the `centre-of-mass'. Since the dispersion is minimised when we
choose $x_{c}$ to be the `centre-of-mass', we can calculate the extent
of the eigenvector directly by

\begin{equation}
D = \min_{x_{c}} R^{1/2}(x_{c}) .
\end{equation}
In practice we do these calculations by decomposing the eigenvector
$|e_{\lambda}\rangle$ into a linear combination of the would-be
zero-modes and working out the dispersions of these using, where
appropriate, some simple approximations.

\begin{figure}[p]
\begin{center}
\leavevmode
\epsfxsize=100mm
\epsfbox{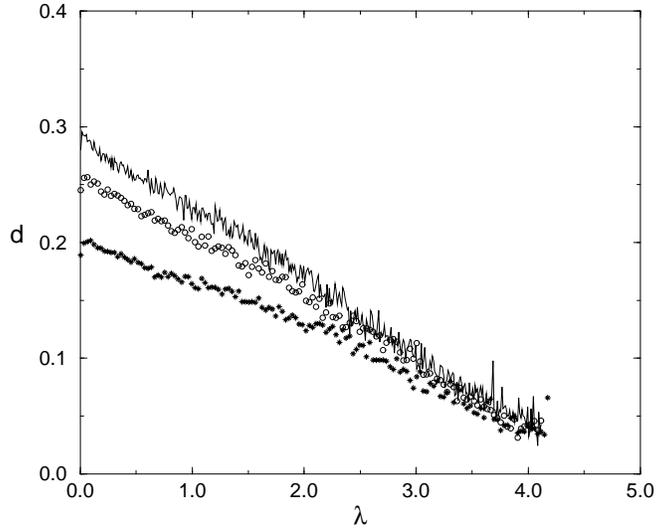}
\caption{Size of eigenfunctions of $\di$  versus $\lambda$. For
volumes  $V=0.4096\ (\star)$, $V=1\ (\circ)$  and 
$V \approx 1.5$ (solid) using $f=1$ hard spheres.}
\label{fig:disp}
\end{center}
\end{figure}

We have performed a calculation with an ensemble of hard-spheres on
volumes $V=0.41, 1.0, 1.5$. The results of $D(\lambda)$ for these
three volumes is shown in figure~\ref{fig:disp}.  We see that the
dispersion increases with decreasing eigenvalue; perhaps roughly
linearly. We also see that there are stronger finite volume effects at
smaller values of $\lambda$. It would appear that the eigenvectors for
small $\lambda$ are being constrained by our finite volumes to be
smaller than they would otherwise be. Whether the dispersion would
actually diverge at $\lambda=0$ in the $V \to \infty$  limit is a
question we obviously cannot answer. Certainly this shows that the
modes that are important for the spontaneous breaking of chiral
symmetry are delocalised on the scale of instanton nearest-neighbour
interactions.  The implications of figure~\ref{fig:disp} are
intriguing and need further investigation.

\section{Full QCD (${\bf n_f}$=1,2)}
\label{ch:unq}

In this section we shall use the simplest hard-sphere version
of our model to analyse QCD with 1 and 2 quark  flavours. 
The calculations are, computationally, far more intensive
than for the pure gauge theory, and we will not be able 
to perform the various checks that we were able to perform
in that case. Thus, although we are reassured by the fact that
this simplest hard-sphere analysis proved remarkably robust
in quenched QCD, we must regard the calculations here as
being of an exploratory nature.

There is a surprisingly large number of non-trivial questions 
we can ask within the framework of our model. We shall focus upon:

\begin{itemize}

\item{Spectral density. How does the spectral density behave with
dynamical quarks~? Do we still get a power divergence as seen
previously, $b(m)\lambda^{-d(m)}$ where $b(m),\ d(m)$ are now
dependent upon the quark mass~? What is the behaviour of the spectral
density as a function of the number of quark flavours $N_{f}$~?}

\item{Chiral condensate. What is the behaviour of \ssi(m)~? 
In particular, do we have spontaneous symmetry breaking and
how does it behave as a function of the number of quark flavours~?}

\item{Topological susceptibility. General arguments give the behaviour
of this quantity as:

\begin{equation}
\begin{array}{lcll}
\langle Q^{2}\rangle & \propto & mV & {\rm symmetry\ broken\ phase} \\
& \propto & m^{N_{f}}V \ \ \ & {\rm symmetric\ phase} .
\end{array}
\label{q2_behav}
\end{equation}
Can our model reproduce such behaviour~?}

\item{Particle masses. In particular, does the mass of the
$\eta^{'}$ meson go to a non-zero limit as the quark mass
$m \to 0$~?}
\end{itemize}

We shall begin by describing in detail how we generate the
instanton configurations for $N_f\not= 0$. We then describe
how we calculate the $\eta^{'}$ and $\sigma$ masses in our
model. From there we move on to our calculations,
in which we address the questions listed above. 

\subsection{Ensemble generation}

The weighting of the instanton ensembles will have a gauge
part and a fermionic part. The gauge part will be the same
as in the quenched case, except that we will allow both
$Q$ and $N$ (the total number of topological charges)
to vary. We vary $Q$ because we want to calculate the 
$m$-dependence of $\langle Q^{2}\rangle$. In addition
we expect that $\langle Q^{2}\rangle$ will cease
to be large once $m$ is small, and then there is
the danger that using only $Q=0$ will bias the physics.
The reason for varying $N$ is less compelling and,
in practice, the main purpose of such calculations
will be as a check on our more extensive fixed-$N$
calculations.

\subsubsection{The gauge weighting}

As in the quenched case, the topological charges are positioned
at random in space-time, with a fixed size ($\rho$=0.2).
When we wish to allow $N=N_{A}+ N_{I}$ to vary, we
choose the number of instantons using a Poisson distribution, 
and the same for anti-instantons. The two distributions are 
taken to be independent:

\begin{equation}
P(N_{A}=s, N_{I}=t) =
\exp(-2\mu)\frac{\mu^{s+t}}{s!t!}
\end{equation}
Note that $\mu$ would be the mean number of instantons (as well as
of anti-instantons) if we used the gauge weighting alone. (This
is similar to equation~\ref{eq:varyia} with $\mu = {\bar{N}}/2$,
albeit with a different variance.)

\subsubsection{The fermion weighting}

In the case when  $N=N_{A}+ N_{I}$ is fixed, the obvious fermion 
weighting to use is one that is calculated entirely within
the basis of would-be zero modes,

\begin{equation}
\det(\di[A] - im) \doteq 
m^{|N_{A}-N_{I}|}
\prod_{i=1}^{\min(N_{I},N_{A})}(\lambda_{i}^{2}+m^{2})
\label{mod-det-Nfix}
\end{equation}
(raised to the power $N_f$) where the $\lambda_{i}$ are the 
eigenvalues we obtain, for the given instanton ensemble, as 
described earlier in this paper. There are of course other
modes but their number does not change, since the total 
number of modes $N_{T}$ is fixed if we are in a fixed volume
(with a fixed ultraviolet cut-off). We have assumed, in
our calculations of the chiral condensate, that
it is a good approximation to neglect the variation of 
these other modes as we alter the instanton gas, and
equation~\ref{mod-det-Nfix} embodies the same assumption.

\begin{figure}[tb]
\begin{center}
\leavevmode
\epsfxsize=100mm
\epsfbox{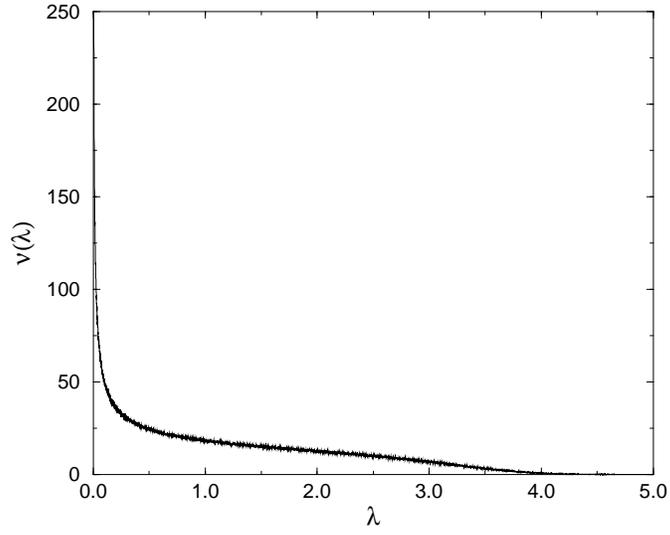}
\end{center}
\caption{The spectral density for $f=1, \ V=1, \ N_{f}=0$ plotted
for the full range of $\lambda$.}
\label{fig:h_stan_lr}
\end{figure}

If, however, we vary  $N=N_{A}+ N_{I}$ then we can no longer 
ignore these other modes: their number, $N_{T}-N$, will vary
and so some choice has to be made for their weighting.
We will make the simplest choice: each mode will be set
to a fixed value $\overline{\lambda}_{NZ}$. In that case
our weighting becomes
\begin{equation}
\det(\di[A] - im) \doteq (\overline{\lambda}_{NZ}^{2} +
m^{2})^{(N_{T}-N_{A}-N_{I})/2}
m^{|N_{A}-N_{I}|}\prod_{i=1}^{\min(N_{I},N_{A})}
(\lambda_{i}^{2}+m^{2})
\label{mod-det}
\end{equation}
(again raised to the power $N_f$). It is clear that we
do not actually have to choose a concrete value for
$N_{T}$ since it is only the ratio of the
determinants that enters into the probability of
changing one configuration of instantons for another one,
and in the ratio the factor 
$(\overline{\lambda}_{NZ}^{2} + m^{2})^{N_{T}/2}$ drops out.
However we do need some choice for the value of
$\overline{\lambda}_{NZ}$. In keeping with our
assumption that the lowest-lying eigenvalues are generated
by the mixing of the would-be zero modes, we shall choose
$\overline{\lambda}_{NZ}$ to be somewhat larger than the
average eigenvalue in the zero-mode basis. A typical
quenched QCD eigenvalue spectrum is shown in 
figure~\ref{fig:h_stan_lr}, and this has an average eigenvalue 
of ${\overline \lambda}\simeq 1.1$; so we shall choose
$\overline{\lambda}_{NZ}=2.0$ in our calculations.
The somewhat arbitrary nature of this choice is 
unsatisfactory and in practice we shall minimise the 
ambiguity by only performing such calculations for 
$m \ll \overline{\lambda}_{NZ}$.
(We remark that in our Monte Carlo we have actually
used $(\overline{\lambda}_{NZ} + m)$ rather than 
$(\overline{\lambda}_{NZ}^{2} + m^{2})^{1/2}$; but 
for $m \ll \overline{\lambda}_{NZ}$ this should not matter.)
Most of our calculations will be for fixed $N$. In this
case the values of $Q$ are necessarily all even or all
odd. This should not introduce a significant bias as long as
$\langle Q^2 \rangle$ is not too small.

\subsubsection{Monte Carlo simulation}

As described above, we incorporate a gauge and fermion
weighting into our model using the two parameters $\mu$ and
$\overline{\lambda}_{NZ}$. The parameter $\mu$ is related to
the packing fraction of our configurations but does not
determine it wholly because the fermion determinant will also
play a part in finding the equilibrium number of objects in the gas.

The Monte Carlo simulation begins with a random configuration. We then
move a single object to generate a new trial configuration. This
process is repeated as we ``sweep'' through the gas, moving each
object in turn, accepting moves according to the standard Metropolis
algorithm:

\begin{equation}
\begin{array}{lcll}
P({\rm accept}) & = & 1 & 
\det({\rm new}) > \det({\rm old})\\
& = & \{{{\det({\rm new})}\over{\det({\rm old})}}\}^{N_f}
 \ \ \ \ & {\rm otherwise} .
\end{array}
\end{equation}
Periodically (usually  every 10 attempted moves) we attempt to either
increase or decrease the number of instantons or anti-instantons by
one. We expect the system
to come into equilibrium after some number of sweeps. As the change
between successive configurations is small (differing only in the
position of a single object or in having one extra or one fewer
object), we clearly require long sequences of configurations,
both to thermalise and to obtain a useful number of effectively 
independent configurations. Whilst we use all the configurations 
that we generate (after equilibriation) our statistical analysis
uses a binned jack-knife procedure that should ensure reasonably
accurate error analyses. 

\begin{figure}[tb]
\begin{center}
\leavevmode
\epsfxsize=100mm
\epsfbox{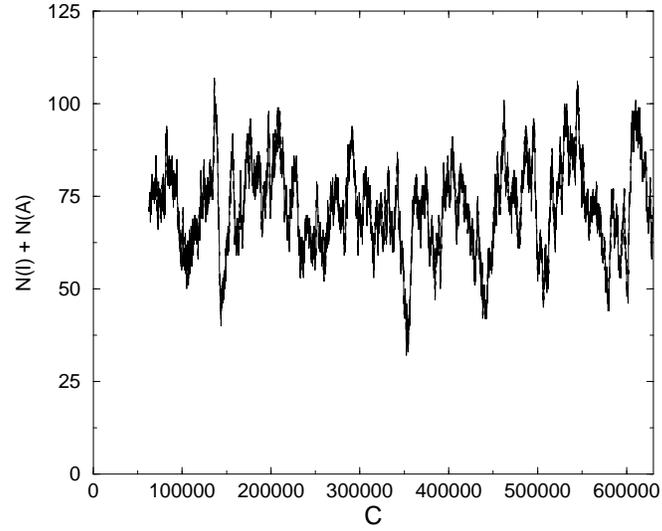}
\end{center}
\caption{$N_{f}=1,\ \overline{\lambda}_{NZ}=2.0,\ V=1,\ m = 0.15$. The
total number of objects as a function of the configuration number
in the Monte Carlo sequence.}
\label{fig:015_eqt_rm1}
\end{figure}

\begin{figure}[tb]
\begin{center}
\leavevmode
\epsfxsize=100mm
\epsfbox{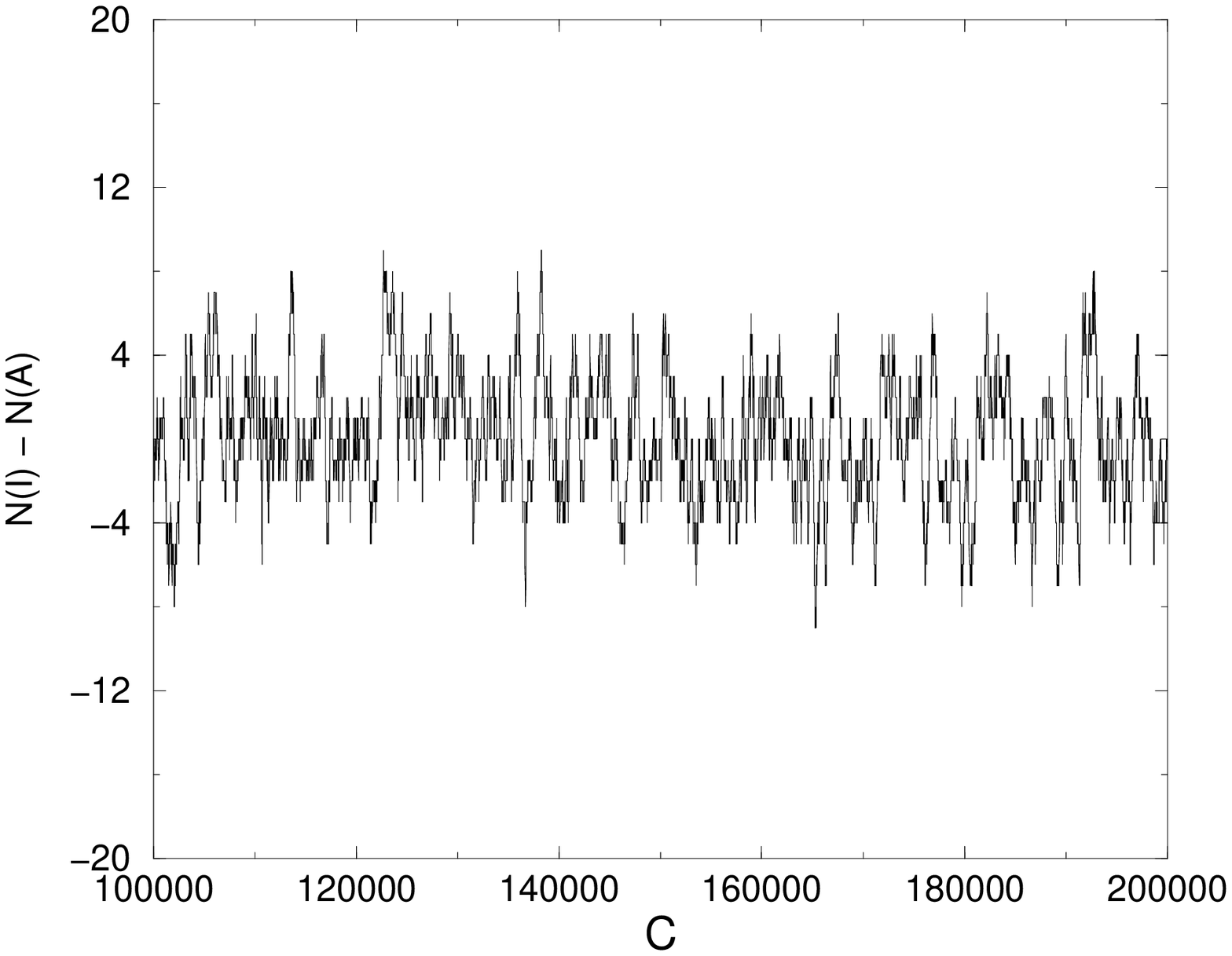}
\end{center}
\caption{$N_{f}=1,\ \overline{\lambda}_{NZ}=2.0,\ V=1,\ m = 0.15$. The
net winding number as a function of the configuration number. We show
only a section of the run for clarity.}
\label{fig:015_eqw_rm1}
\end{figure}

\begin{figure}[tb]
\begin{center}
\leavevmode
\epsfxsize=100mm
\epsfbox{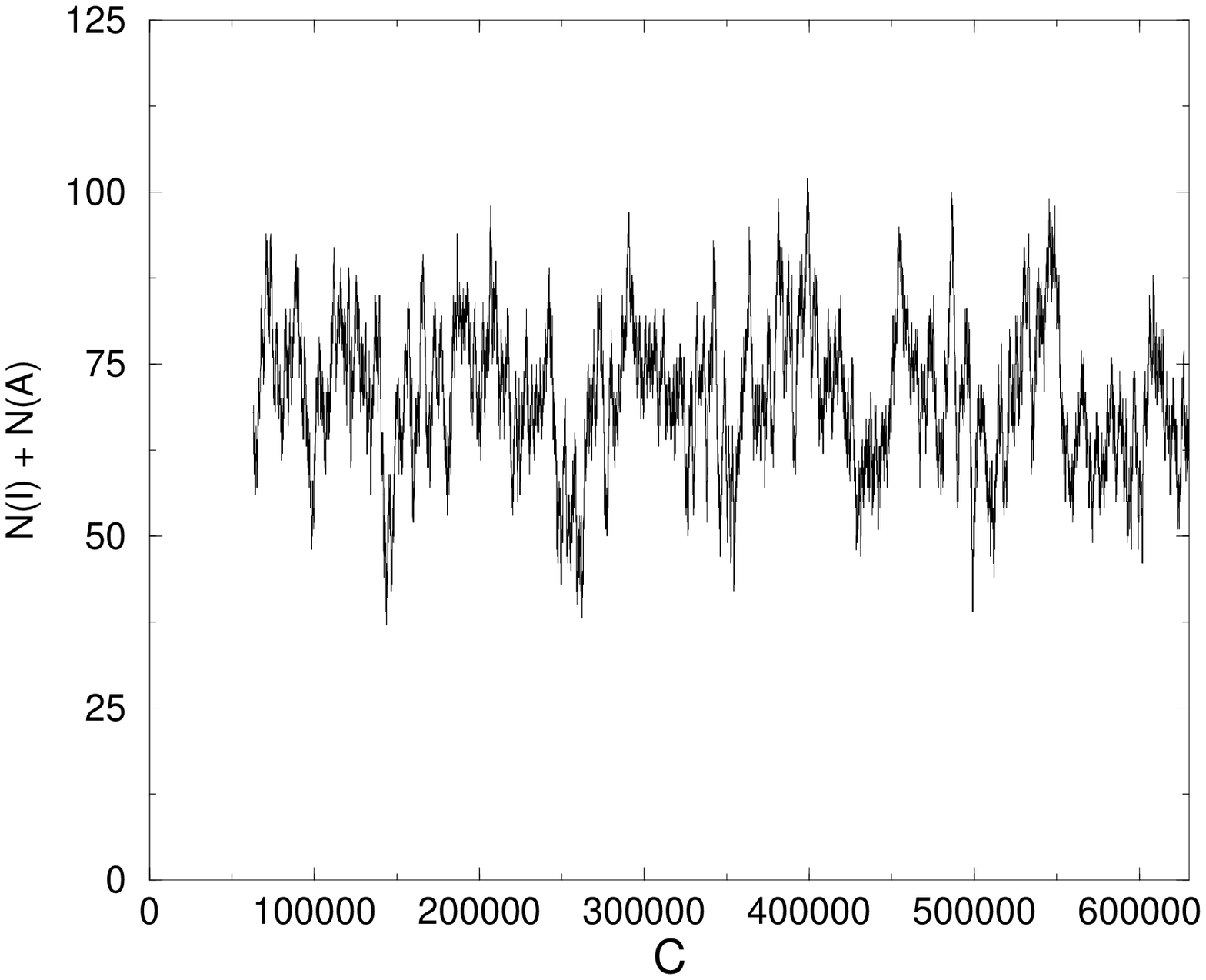}
\end{center}
\caption{$N_{f}=1,\ \overline{\lambda}_{NZ}=2.0,\ V=1,\ m = 0.5$. The
total number of objects as a function of the configuration number.}
\label{fig:050_eqt_rm1}
\end{figure}

\begin{figure}[tb]
\begin{center}
\leavevmode
\epsfxsize=100mm
\epsfbox{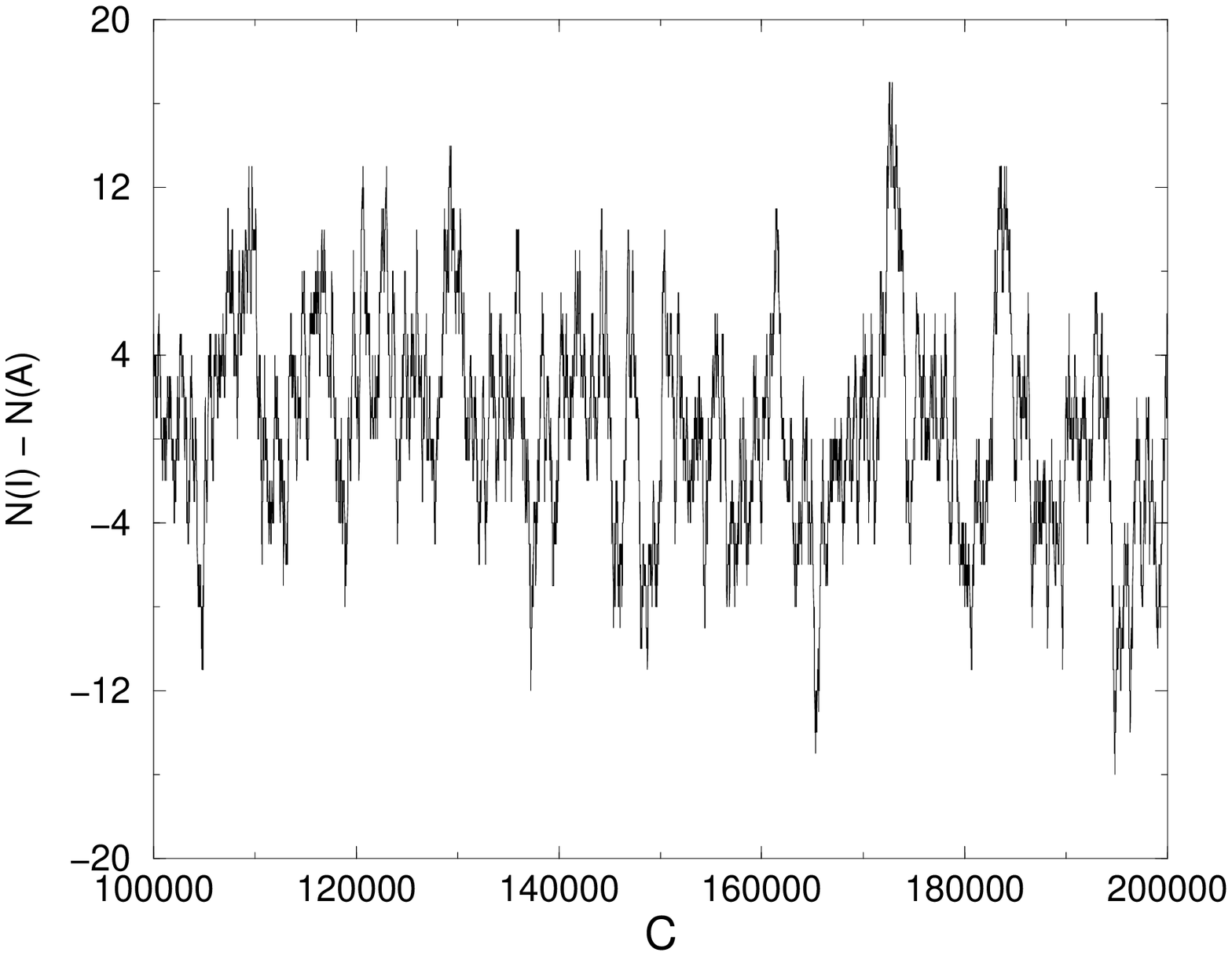}
\end{center}
\caption{$N_{f}=1,\ \overline{\lambda}_{NZ}=2.0,\ V=1,\ m = 0.5$. The
net winding number as a function of the configuration number. We show
only a section of the run for clarity.}
\label{fig:050_eqw_rm1}
\end{figure}

To ensure thermalisation we monitor the values taken by
$N = N_I + N_A$ and $Q = N_I - N_A$ during the sequence 
of instanton configurations generated by the Monte Carlo.
In practice we find that a sequence of 630000
configurations appears to be sufficient to explore the
available phase space; and that after 63000 configurations
we appear to have thermalised (i.e. lost all memory of the
random starting configuration). This is illustrated in
figure~\ref{fig:015_eqt_rm1} and figure~\ref{fig:015_eqw_rm1}
for the lightest quark mass we use, and in 
figure~\ref{fig:050_eqt_rm1} and figure~\ref{fig:050_eqw_rm1}
for an intermediate quark mass. As one would expect the
phase space is explored more weakly for the lighter mass.

\subsection{Calculating masses}

The usual way to calculate masses in a (Euclidean)
field theory is from the $t$-dependence of appropriate
correlation functions. The latter can be estimated
by standard Monte carlo techniques.  We can attempt to
apply the same approach to our instanton ensembles.
Since we calculate the eigenvalues (and can easily
calculate the corresponding eigenvectors) of $\di$, we can, 
in principle, calculate quark and hence hadron propagators.
Here, however, we shall limit ourselves to calculations
that involve only `gluonic' operators. That is to say,
we calculate the masses of flavour-singlet mesons.
This includes what is perhaps the most interesting
meson in this context, the $\eta^\prime$. We shall also 
calculate the mass of the scalar $\sigma$. 

The $\eta^\prime$ has quantum numbers $0^{-+}$ which
are also the quantum numbers of the topological charge
density. Thus it can be calculated from the
correlation function, $\langle Q(0)Q(t)\rangle$,
assuming the usual decomposition

\begin{equation}
\langle Q(0)Q(t)\rangle = \sum_{n}c_{n}\exp(-M_{n}t) .
\label{eqn_Q_cor}
\end{equation}
The lowest mass should be the $\eta^\prime$. In 
practice what we do is to divide up our volume into a 
number of ``strips'' each of width $\delta t \ll 1$.
We calculate the total `charge' in a given strip simply by
adding up the charges of all the objects whose centres are 
contained within the strip. We call this $Q(t)$ for the 
strip $[t,t+\delta t)$. That is to say, $Q(t) = N_{I}(t) - N_{A}(t)$ 
where $N_{I/A}(t)$ are the number of instantons/anti-instantons 
whose centres are contained within the strip. This differs
from the actual topological charge density within a strip,
but is clearly an equally good operator. (The difference is due 
to the spread of the instanton core; but that is modelled by some
relatively arbitrary function within our model and so there is no
point in including that.)

One might worry that the lightest mass contributing to 
the correlator in equation~\ref{eqn_Q_cor} might be a  
$0^{-+}$ glueball rather than the $\eta^\prime$. 
However we note that in the pure gauge theory the
topological charges are completely uncorrelated,
so the correlation length is zero, i.e. the
$0^{-+}$ glueball mass is $\infty$. Thus it seems
safe to assume that it can be ignored in full QCD
as well.

The $\sigma$ has quantum numbers $0^{++}$ and so can be 
calculated from correlators of the gluonic action density, 
$\langle S_g(0)S_g(t)\rangle$. Within our model the gluonic 
action density is simply proportional to the density of 
topological charges and so we use the correlator of
$N(t) = N_{I}(t) + N_{A}(t)$, i.e. the total number of charges
whose centres are located within the strip labelled by $t$. 
For the same reason as in the case of the $0^{-+}$, we can
ignore the glueballs. However there is now a non-trivial
vacuum state contribution, and so we have to use the
vacuum subtracted correlator,
$\langle N(0)N(t)\rangle - \langle N(0)\rangle^{2}$.
The large-$t$ exponential fall-off of this correlator
should then give us the mass of the $\sigma$ meson.
(Note that if we fix $N$ in our calculation, this
induces long-range correlations that are not due
to particle exchange. For this reason we do not
calculate the $\sigma$ mass in our fixed-$N$ calculations.)

To make sure that one has identified the large-$t$
exponential decay of a correlation function, it is
useful to define an effective mass

\begin{equation}
m_{eff}(t) = -\ln\left(\frac{\langle
{\cal O}(0){\cal O}(t)\rangle}{\langle{\cal O}(0){\cal O}(t-1)\rangle}\right)
,
\label{eqn_meff}
\end{equation}
where we have written our correlator with a generic operator ${\cal
O}$ which can represent either the charge or the number density. It
should be apparent that if the correlation function is in fact given
by a simple  exponential for $t \ge t_0$ then our effective 
mass $m_{eff}(t)$ will be independent of $t$, within errors,
for $t \ge t_0$. If $t_{min}$ is the smallest value of $t_0$
for which this is the case, then $m_{eff}(t_{min})$ provides
us with an estimate of the lightest mass, and its error.

It is of course the case that our model is not a field theory
and a decomposition of the kind given in equation~\ref{eqn_Q_cor}
may not hold. However, since our model is intended to provide
an approximation to the relevant infrared physics of QCD,
we will take the masses we calculate in this way as being
analogues, within our model, of the corresponding masses
in QCD.

\subsection{Calculations}

We perform calculations for both 1 and 2 quark flavours.
For $N_f=1$ one expects no Goldstone boson since the one
would-be Goldstone boson, the $\eta^\prime$, acquires a mass
through the anomaly. (For simplicity we shall 
refer to the flavour singlet pseudoscalar as the $\eta^\prime$ 
even though that is, strictly speaking, the flavour singlet
member of the nonet with $N_f=3$.) Although the $U(1)$ chiral
symmetry is broken by the anomaly, it is still interesting
to see whether the Dirac spectrum possesses a non-zero
$\lambda \to 0$ limit. The $N_f =2$ case, on the other hand,
contains essentially all the features of the physically
relevant $N_f =2+1$ case.
 
We will carry out calculations separately
for $N$ fixed and for $N$ being
allowed to vary. The latter calculations will show that 
$\langle N \rangle$ varies weakly with the quark mass $m$;
in contrast to $\langle Q^2 \rangle$. This suggests that
constraining $N$ to be fixed is not a serious bias.
Most of our calculations will in fact be for $N$ fixed,
with the variable $N$ calculations serving as a check.
Fixing $N$ has the advantage that we can avoid our
rather ad hoc representation of the non-zero modes 
through an average eigenvalue, $\overline{\lambda}_{NZ}$.
In addition we can see from 
figures~\ref{fig:015_eqt_rm1}-~\ref{fig:050_eqw_rm1}
that the Monte Carlo explores the fluctuations in $N$
much more slowly than the  fluctuations in $Q$. Thus a
fixed-$N$ calculation may also prove to be statistically
more accurate.

All the calculations described here will be with hard-sphere
wavefunctions with radius $\rho=0.2$ and in a space-time
volume $V=1$. Our fixed $N$ calculations will contain
$N=126$ which corresponds to a packing fraction $f=1.0$.
When we vary $N$ we choose the gauge weighting so that the 
average value of $N$ would be 126 if there were no fermionic
weighting. In practice the latter reduces $\langle N \rangle$ 
so that the packing fraction lies in the interval
$f \in [0.35,0.55]$. 

For the calculations at fixed $N$, we choose quark masses
$m = 0.15,\ 0.3,\ 0.5,\ 1.0,\ 2.0$ and $3.0$. For the
calculations with varying $N$ we only use the lowest
3 values of $m$. This is because the fermionic weighting
will presumably become sensitive to the precise value of
$\overline{\lambda}_{NZ}$ once $m$ becomes comparable to it.

We expect that in our finite volume, we will lose chiral
symmetry breaking once $m$ becomes too small. To gain some
intuition as to when that might occur, it is useful to
express everything in physical `MeV' units. If we take the
instanton radius to be about $0.5fm$ \cite{DSMT} then
the length of our space-time, being $5\times\rho$, is
about $2.5fm \simeq 1/80MeV$. So $m=1$ corresponds to
$m=80 MeV$, i.e. just a little lighter than the strange
quark mass, and $m=0.15$ corresponds to $m\simeq 12 MeV$, 
i.e. close to the physical $u,\ d$ masses. Thus we should 
certainly not be surprised to see finite-$V$ effects at 
the lower end of our range of masses.

We now turn to the detailed results of our calculations,
the parameters of which are summarised in table~\ref{tab:unq}.

\subsubsection{$N_{f} = 1$; variable $N$.}

\begin{figure}[tb]
\begin{center}
\leavevmode
\epsfxsize=100mm
\epsfbox{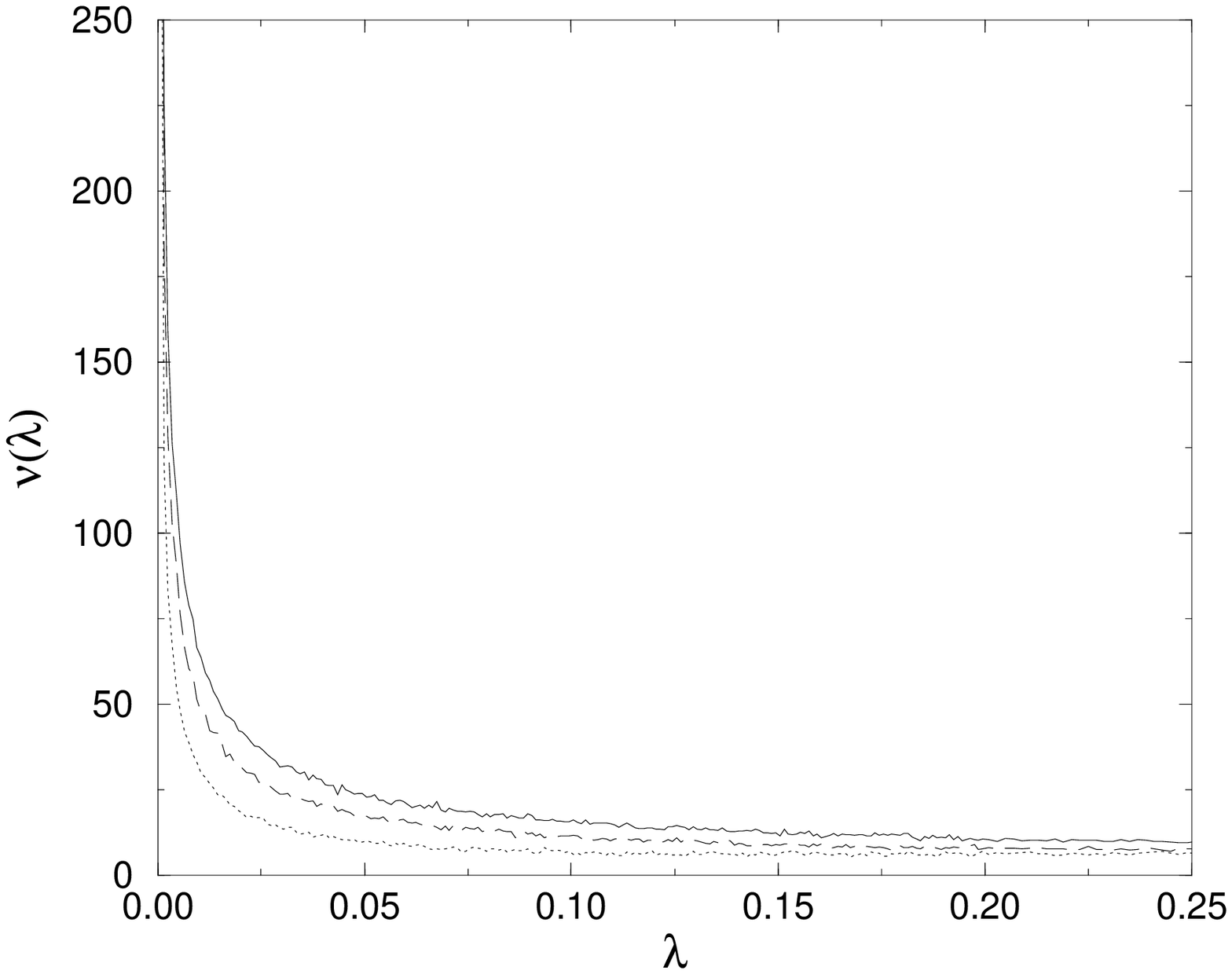}
\end{center}
\caption{$N_{f}=1,\ \overline{\lambda}_{NZ}=2.0,\ V=1$. The spectral
density for quark mass $m=0.5$ (solid), 0.3 (long dashed) and 0.15
(dotted).}
\label{fig:sp_un_rm1}
\end{figure}

\begin{figure}[tb]
\begin{center}
\leavevmode
\epsfxsize=100mm
\epsfbox{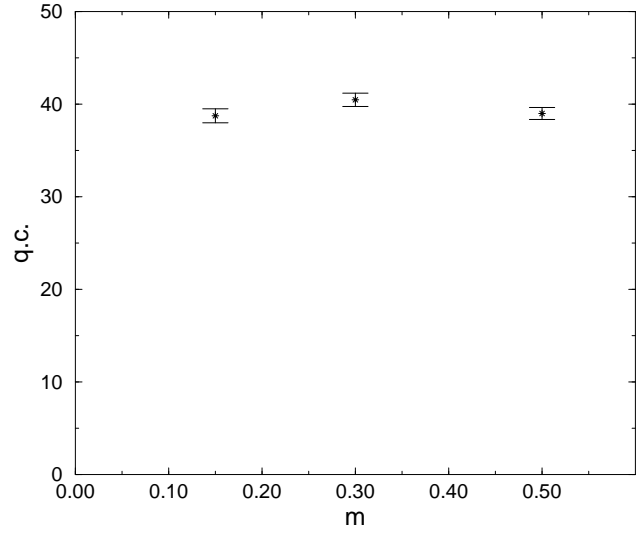}
\end{center}
\caption{The quark condensate \ssi\ as a function of the quark mass
$m$ from the spectra plotted in figure~\ref{fig:sp_un_rm1}.}
\label{fig:qc_rm1}
\end{figure}

In figure~\ref{fig:sp_un_rm1} we display the spectral density
of the Dirac operator for quark masses, $m = 0.15,\ 0.3,\ 0.5$.
The first thing we note is that the spectral density appears 
to diverge as $\lambda \rightarrow 0$ for all three masses.
It is however, equally clear that the spectral density is not
independent of the quark mass. Indeed, if we fit the spectra
to the power-law form in equation~\ref{eqn-power-law} then
we find, as shown in table~\ref{tab:sp}, that the coefficient 
of the divergence appears to vanish, $b \rightarrow 0$, as
$m \rightarrow 0$. Thus the divergence in the spectral density
need not lead to a divergent quark condensate of the kind that 
we encountered in quenched QCD. This appears to be confirmed
by a direct calculation of the chiral condensate against $m$, as
shown in figure~\ref{fig:qc_rm1}, where we see that we appear to 
have chiral symmetry breaking with a finite condensate.

\begin{figure}[tb]
\begin{center}
\leavevmode
\epsfxsize=100mm
\epsfbox{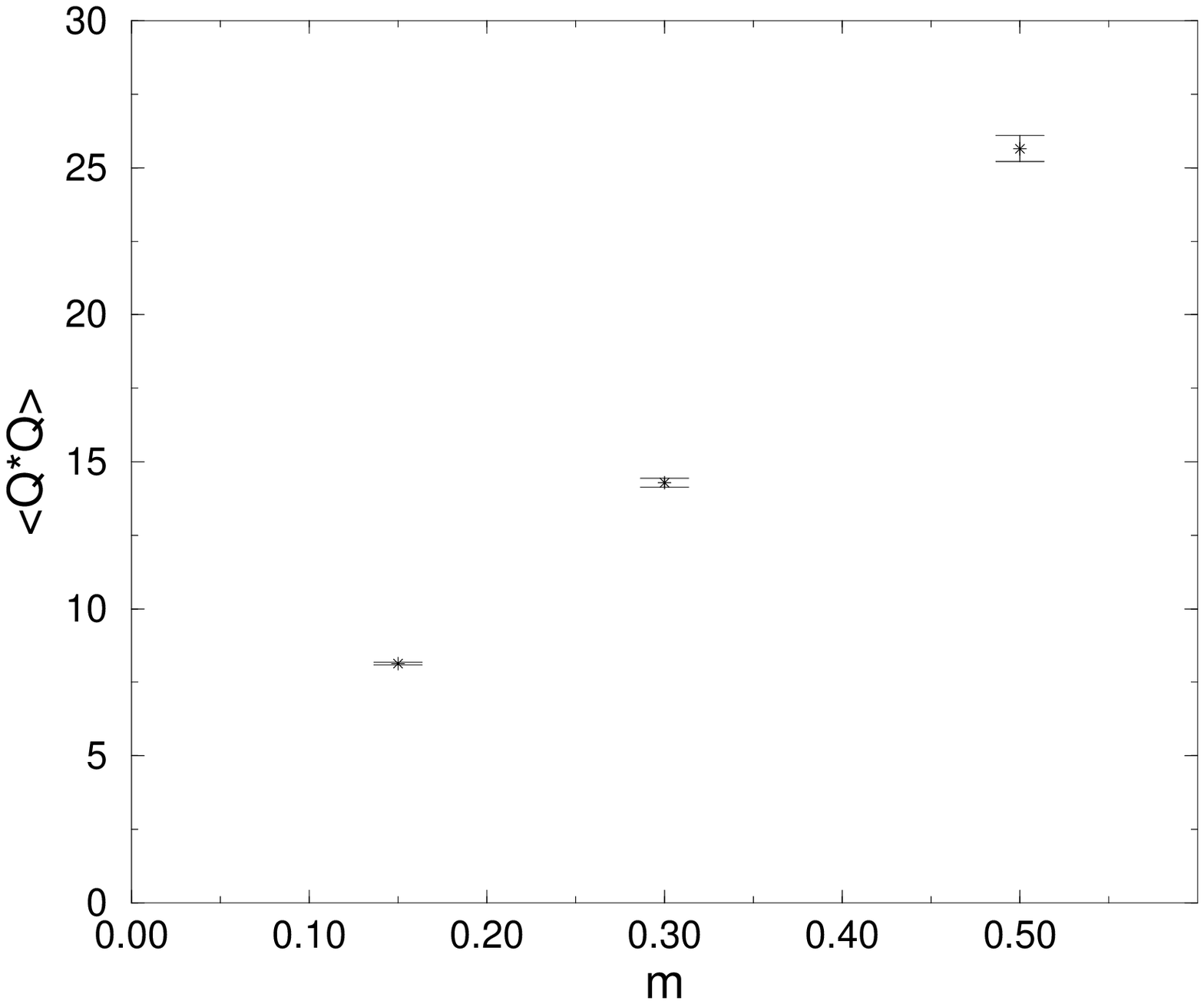}
\end{center}
\caption{$N_{f}=1,\ \overline{\lambda}_{NZ}=2.0,\ V=1$. The second
moment of the winding number distribution $\langle Q^{2}\rangle$ as a
function of the quark mass $m$.}
\label{fig:qq_rm1}
\end{figure}

We also calculate $\langle Q^{2}\rangle$ as a function
of $m$, and list the values in table~\ref{tab:unq}.
As we see from figure~\ref{fig:qq_rm1} the variation
of  $\langle Q^{2}\rangle$ is approximately linear
with $m$ in accord with what we expect from the anomalous 
Ward identities for QCD: 
\begin{equation}
{{\langle Q^{2}\rangle} \over {V}}
=
{{m \ssi} \over {N^2_f}}
+
O(m^2).
\label{chi_ward}
\end{equation}
Indeed, if we insert into this equation the values relevant to
our simulations, i.e. $V=1$, $N_f=1$ and the value for $\ssi$ 
shown in figure~\ref{fig:qc_rm1}, then we obtain predicted values 
for $\langle Q^{2}\rangle$ that are within $\sim$25\% of 
the values shown in figure~\ref{fig:qq_rm1}. This is quite
remarkable.

\begin{figure}[tb]
\begin{center}
\leavevmode
\epsfxsize=100mm
\epsfbox{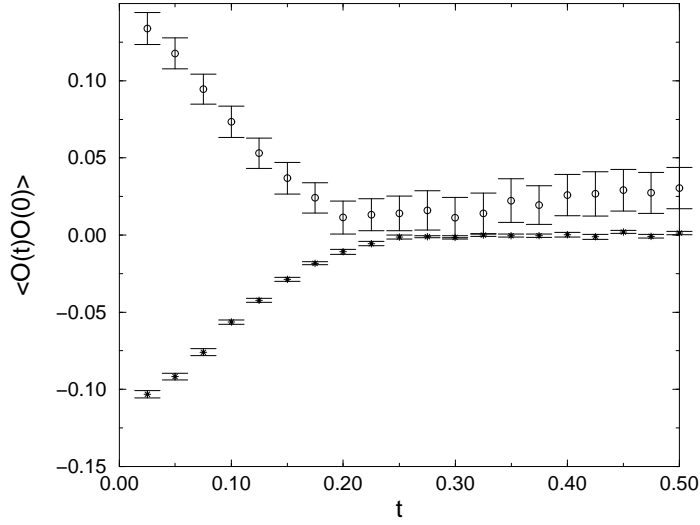}
\end{center}
\caption{$N_{f}=1,\ \overline{\lambda}_{NZ}=2.0,\ V=1, \ m=0.15$. 
Correlation functions for $\eta^{'}\ (\star)$, and for $\sigma\ (\circ)$.}
\label{eta_sig_corr}
\end{figure}

\begin{figure}[tb]
\begin{center}
\leavevmode
\epsfxsize=100mm
\epsfbox{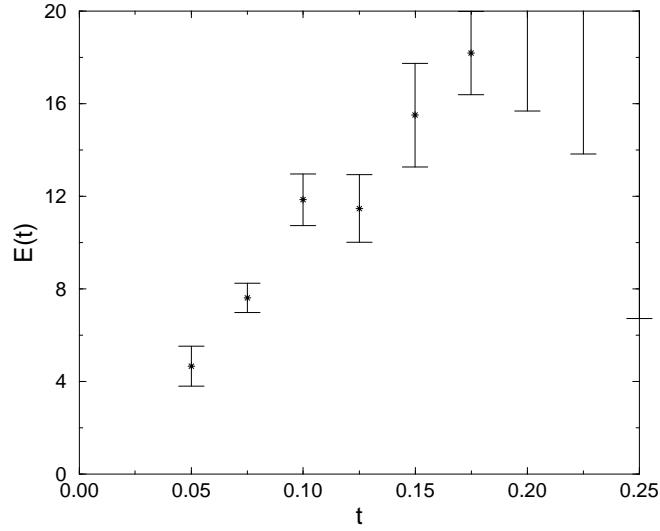}
\end{center}
\caption{$N_{f}=1,\ \overline{\lambda}_{NZ}=2.0,\ V=1,\ m=0.15$. The
$\eta^{'}$ effective mass from the correlation functions shown in
figure~\ref{eta_sig_corr}.}
\label{meff_eta}
\end{figure}

We now turn to our calculation of the flavour singlet meson masses. 
In figure~\ref{eta_sig_corr} we show an example of the correlation 
functions from which we extract the $\eta^{'}$ and $\sigma$ masses.
We can use equation~\ref{eqn_meff} to extract effective masses,
as shown for the $\eta^{'}$ in figure~\ref{meff_eta}. The
errors are large and there is not really much evidence that 
$m_{eff}(t)$ plateaus at large $t$ to a finite mass,  
$m_{\eta^{'}}$. In this particular case we choose the
effective mass at $t=0.1$ as our estimate of $m_{\eta^{'}}$
since, roughly speaking, effective masses at larger values of
$t$ are consistent with it within errors. We list in 
table~\ref{tab:sp} our mass estimates, obtained in this way.
Despite the crudity of our calculation 
it seems clear that $m_{\eta^{'}}$ does not vanish
as $m \to 0$ : the ${\eta^{'}}$ is not a Goldstone boson.
In our units, where $\rho=0.2$, we find $m_{\eta^{'}}\sim 12$.
If we transform to `MeV' units, using  $\rho\sim 0.5fm$, then
this mass becomes $m_{\eta^{'}}\sim 1GeV$. In QCD this mass
is generated by topological fluctuations, and through
large-$N_c$ arguments can be related to the value of
$\langle Q^{2}\rangle$ in the quenched theory. With our
gauge weighting we would obtain in the quenched 
theory $\langle Q^{2}\rangle = N = 126$ and hence, in
`MeV' units, $\langle Q^{2}\rangle/V \sim (260 MeV)^4$.
This is not so far from the true value\cite{DSMT} in quenched 
QCD, $\sim (200 MeV)^4$, and shows that our model is
doing remarkably well in reproducing the link between
topology and physics in full QCD.
 
In table~\ref{tab:sp} we also list our estimates of
the mass of the $\sigma$. It is comparable in mass
to the  ${\eta^{'}}$, although perhaps slightly lighter.

\subsubsection{$N_{f} = 1$; fixed $N$.}

\begin{figure}[tb]
\begin{center}
\leavevmode
\epsfxsize=80mm
\epsfbox{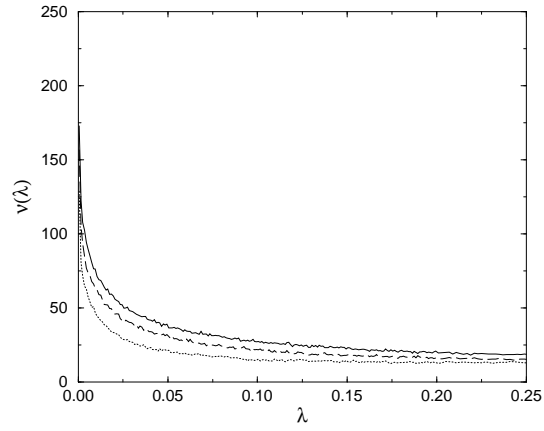}
\end{center}
\caption{$N_{f}=1$, Fixed $N,\ V=1$. The spectral
density for quark mass $m=0.5$ (solid), 0.3 (long dashed) and 0.15
(dotted).}
\label{fig:sp_un_rx1}
\end{figure}

\begin{figure}[tb]
\begin{center}
\leavevmode
\epsfxsize=80mm
\epsfbox{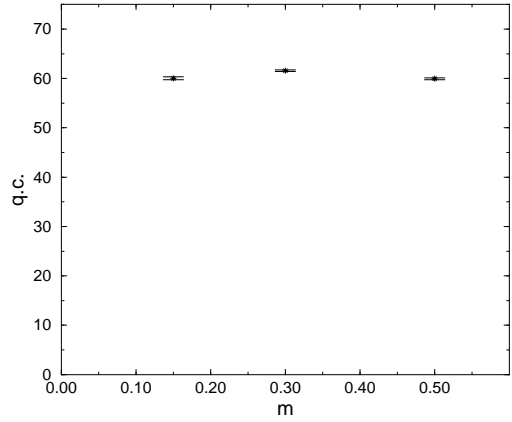}
\end{center}
\caption{$N_{f}=1$, Fixed $N,\ V=1$. The quark condensate \ssi\ as
a function of the quark mass $m$.}
\label{fig:qc_rx1}
\end{figure}

\begin	{figure}[p]
\begin	{center}
\leavevmode
% GNUPLOT: LaTeX picture
\setlength{\unitlength}{0.240900pt}
\ifx\plotpoint\undefined\newsavebox{\plotpoint}\fi
\sbox{\plotpoint}{\rule[-0.200pt]{0.400pt}{0.400pt}}%
\begin{picture}(1500,900)(0,0)
\font\gnuplot=cmr10 at 12pt
\gnuplot
\sbox{\plotpoint}{\rule[-0.200pt]{0.400pt}{0.400pt}}%
\put(120.0,31.0){\rule[-0.200pt]{4.818pt}{0.400pt}}
\put(108,31){\makebox(0,0)[r]{{$0$}}}
\put(1436.0,31.0){\rule[-0.200pt]{4.818pt}{0.400pt}}
\put(120.0,193.0){\rule[-0.200pt]{4.818pt}{0.400pt}}
\put(108,193){\makebox(0,0)[r]{{$3$}}}
\put(1436.0,193.0){\rule[-0.200pt]{4.818pt}{0.400pt}}
\put(120.0,354.0){\rule[-0.200pt]{4.818pt}{0.400pt}}
\put(108,354){\makebox(0,0)[r]{{$6$}}}
\put(1436.0,354.0){\rule[-0.200pt]{4.818pt}{0.400pt}}
\put(120.0,516.0){\rule[-0.200pt]{4.818pt}{0.400pt}}
\put(108,516){\makebox(0,0)[r]{{$9$}}}
\put(1436.0,516.0){\rule[-0.200pt]{4.818pt}{0.400pt}}
\put(120.0,678.0){\rule[-0.200pt]{4.818pt}{0.400pt}}
\put(108,678){\makebox(0,0)[r]{{$12$}}}
\put(1436.0,678.0){\rule[-0.200pt]{4.818pt}{0.400pt}}
\put(120.0,839.0){\rule[-0.200pt]{4.818pt}{0.400pt}}
\put(108,839){\makebox(0,0)[r]{{$15$}}}
\put(1436.0,839.0){\rule[-0.200pt]{4.818pt}{0.400pt}}
\put(120.0,31.0){\rule[-0.200pt]{0.400pt}{4.818pt}}
\put(120,19){\makebox(0,0){\shortstack{\\ \\ \\ {$0$}}}}
\put(120.0,873.0){\rule[-0.200pt]{0.400pt}{4.818pt}}
\put(322.0,31.0){\rule[-0.200pt]{0.400pt}{4.818pt}}
\put(322,19){\makebox(0,0){\shortstack{\\ \\ \\ {$0.5$}}}}
\put(322.0,873.0){\rule[-0.200pt]{0.400pt}{4.818pt}}
\put(525.0,31.0){\rule[-0.200pt]{0.400pt}{4.818pt}}
\put(525,19){\makebox(0,0){\shortstack{\\ \\ \\ {$1$}}}}
\put(525.0,873.0){\rule[-0.200pt]{0.400pt}{4.818pt}}
\put(727.0,31.0){\rule[-0.200pt]{0.400pt}{4.818pt}}
\put(727,19){\makebox(0,0){\shortstack{\\ \\ \\ {$1.5$}}}}
\put(727.0,873.0){\rule[-0.200pt]{0.400pt}{4.818pt}}
\put(930.0,31.0){\rule[-0.200pt]{0.400pt}{4.818pt}}
\put(930,19){\makebox(0,0){\shortstack{\\ \\ \\ {$2$}}}}
\put(930.0,873.0){\rule[-0.200pt]{0.400pt}{4.818pt}}
\put(1132.0,31.0){\rule[-0.200pt]{0.400pt}{4.818pt}}
\put(1132,19){\makebox(0,0){\shortstack{\\ \\ \\ {$2.5$}}}}
\put(1132.0,873.0){\rule[-0.200pt]{0.400pt}{4.818pt}}
\put(1335.0,31.0){\rule[-0.200pt]{0.400pt}{4.818pt}}
\put(1335,19){\makebox(0,0){\shortstack{\\ \\ \\ {$3$}}}}
\put(1335.0,873.0){\rule[-0.200pt]{0.400pt}{4.818pt}}
\put(120.0,31.0){\rule[-0.200pt]{321.842pt}{0.400pt}}
\put(1456.0,31.0){\rule[-0.200pt]{0.400pt}{207.656pt}}
\put(120.0,893.0){\rule[-0.200pt]{321.842pt}{0.400pt}}
\put(-48,558){\makebox(0,0){{\large{$\rm{m_{\eta^{\prime}}}$}}}}
\put(788,-53){\makebox(0,0){{\large{m}}}}
\put(120.0,31.0){\rule[-0.200pt]{0.400pt}{207.656pt}}
\put(1335,510){\circle*{12}}
\put(930,489){\circle*{12}}
\put(525,651){\circle*{12}}
\put(322,661){\circle*{12}}
\put(241,699){\circle*{12}}
\put(181,683){\circle*{12}}
\put(1335.0,343.0){\rule[-0.200pt]{0.400pt}{80.701pt}}
\put(1325.0,343.0){\rule[-0.200pt]{4.818pt}{0.400pt}}
\put(1325.0,678.0){\rule[-0.200pt]{4.818pt}{0.400pt}}
\put(930.0,354.0){\rule[-0.200pt]{0.400pt}{65.043pt}}
\put(920.0,354.0){\rule[-0.200pt]{4.818pt}{0.400pt}}
\put(920.0,624.0){\rule[-0.200pt]{4.818pt}{0.400pt}}
\put(525.0,532.0){\rule[-0.200pt]{0.400pt}{57.093pt}}
\put(515.0,532.0){\rule[-0.200pt]{4.818pt}{0.400pt}}
\put(515.0,769.0){\rule[-0.200pt]{4.818pt}{0.400pt}}
\put(322.0,548.0){\rule[-0.200pt]{0.400pt}{54.443pt}}
\put(312.0,548.0){\rule[-0.200pt]{4.818pt}{0.400pt}}
\put(312.0,774.0){\rule[-0.200pt]{4.818pt}{0.400pt}}
\put(241.0,543.0){\rule[-0.200pt]{0.400pt}{75.161pt}}
\put(231.0,543.0){\rule[-0.200pt]{4.818pt}{0.400pt}}
\put(231.0,855.0){\rule[-0.200pt]{4.818pt}{0.400pt}}
\put(181.0,559.0){\rule[-0.200pt]{0.400pt}{59.743pt}}
\put(171.0,559.0){\rule[-0.200pt]{4.818pt}{0.400pt}}
\put(171.0,807.0){\rule[-0.200pt]{4.818pt}{0.400pt}}
\end{picture}

\end	{center}
\vskip 0.15in
\caption{$N_{f}=1,\ V=1$. The mass of the  $\eta^{'}$ as a function 
of the quark mass.}
\label{eta_sig_mass}
\end 	{figure}
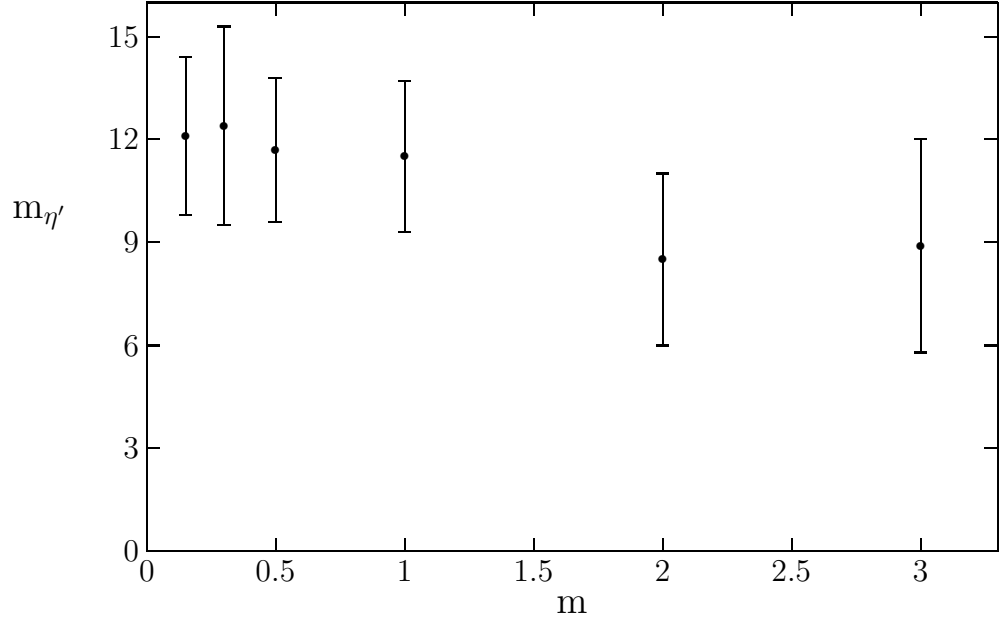

The results of our calculations at fixed $N$ are summarised
in tables~\ref{tab:unq} and~\ref{tab:sp}. Since the packing
fraction is larger than for the variable-$N$ calculations
(126 topological charges per unit volume versus $\sim$70)
we do not expect agreement even in the common $m$-range.
However the qualitative agreement is quite satisfactory.
In  figure~\ref{fig:sp_un_rx1} we plot the spectral density
in this mass range and we see the same features as we saw in
figure~\ref{fig:sp_un_rm1}. The chiral condensate is plotted 
in figure~\ref{fig:qc_rx1}. We see that there is a finite
condensate as $m \to 0$,  but that it is larger than the
one shown in  figure~\ref{fig:qc_rm1}. However the factor 
by which it is larger is close to the trivial factor of
$\sim 126/70$ (the ratio of the number of charges). 
As for  $\langle Q^{2}\rangle$, we see an
approximate linear decrease for $m \leq 1.0$. The actual 
value is larger than for the variable-$N$ case although
by less than the factor of $\sim 126/70$. (One would only 
expect precisely this factor in the dilute gas limit.)
For larger values of $m$ the linear increase must flatten
off, as it does, because it must approach the quenched
($m \to \infty$) value of $\langle Q^{2}\rangle = N = 126$.
The calculated values of the ${\eta^{'}}$ mass, as
shown in figure~\ref{eta_sig_mass}, are consistent with 
those obtained previously (and confirm that the  ${\eta^{'}}$
is indeed not a Goldstone boson). All this
confirms our expectation that fixing $N$ does not
alter the physics of the model in any significant way.

The larger range of $m$ gives us a clearer perspective
on the $\propto b/\lambda^d$ divergence in the spectral
density. We see from table~\ref{tab:sp} 
that the power $d$ is independent of $m$
except at the very smallest value of $m$. We provisionally
ascribe this to a finite volume effect. (The reason for doing
so will become clearer once we present the $N_f =2$ 
calculations.) We note that the value of $d$ is smaller 
than the one we obtained for the variable-$N$ calculation.
Since the packing fraction of the latter is smaller, this
mirrors what we found in quenched QCD: the divergence
weakens with increasing instanton density. Although
$d$ appears to be independent of $m$, this is clearly not 
the case for the coefficient $b$. In fact, in the
intermediate mass range, $m \in [0.3,1.0]$, the dependence 
of $b$ is consistent, within errors, with
$b(m)/\lambda^{d(m)} = b_0 (m/\lambda)^{d_0}$
as suggested by the simple argument leading
to eqn~\ref{div-qcd}.

\subsubsection{$N_{f} = 2$; variable $N$.}

The spontaneous breaking of chiral symmetry that we found
for $N_f = 1$ is academic since the symmetry is in any
case anomalous. We now turn to the corresponding $N_{f} = 2$
calculations where there is a non-anomalous part of the
chiral symmetry that may be spontaneously broken.

\begin{figure}[tb]
\begin{center}
\leavevmode
\epsfxsize=100mm
\epsfbox{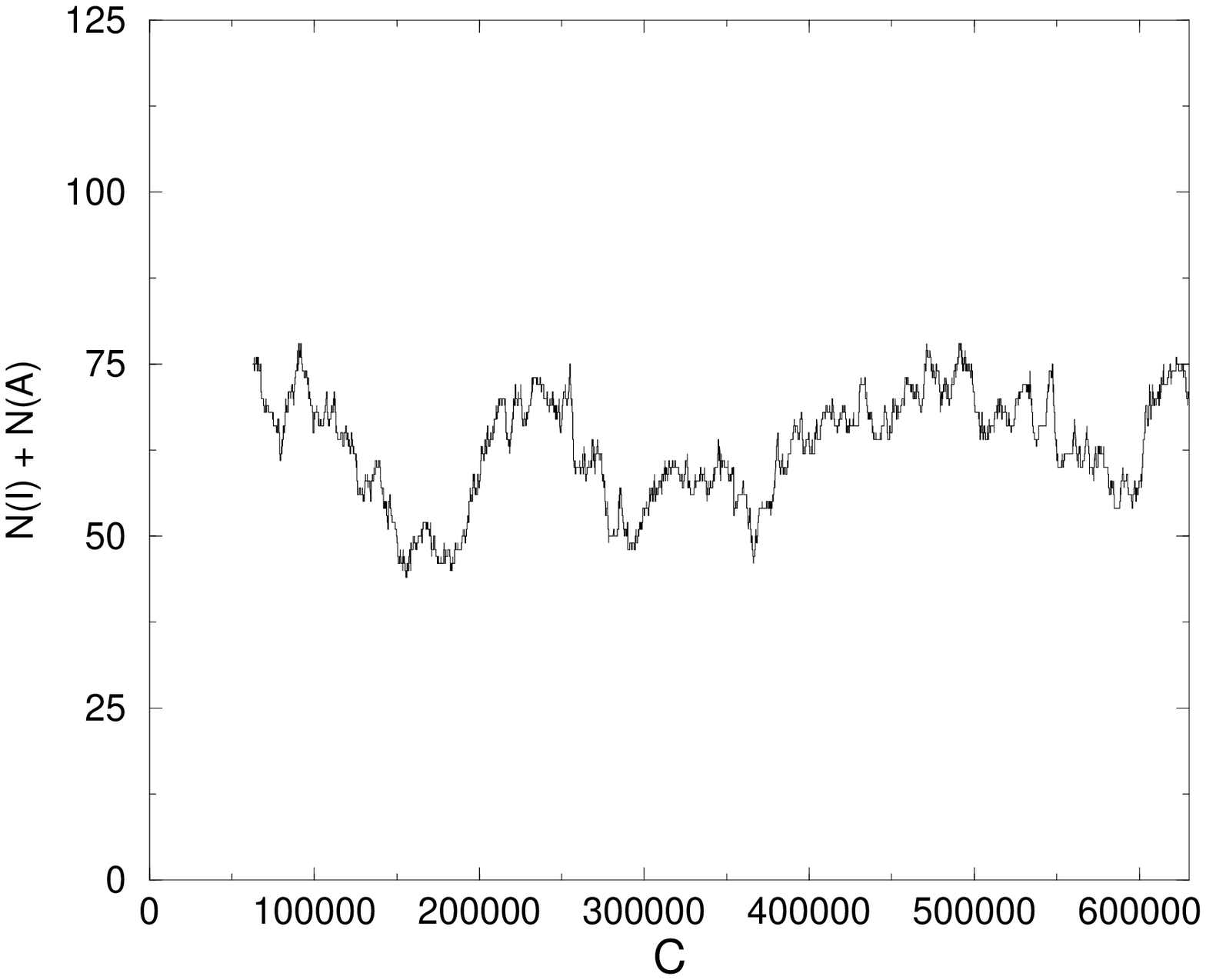}
\end{center}
\caption{$N_{f}=2,\ \overline{\lambda}_{NZ}=2.0,\ V=1,\ m = 0.15$. The
total number of objects as a function of the configuration number.}
\label{fig:015_eqt_rm2}
\end{figure}

\begin{figure}[tb]
\begin{center}
\leavevmode
\epsfxsize=100mm
\epsfbox{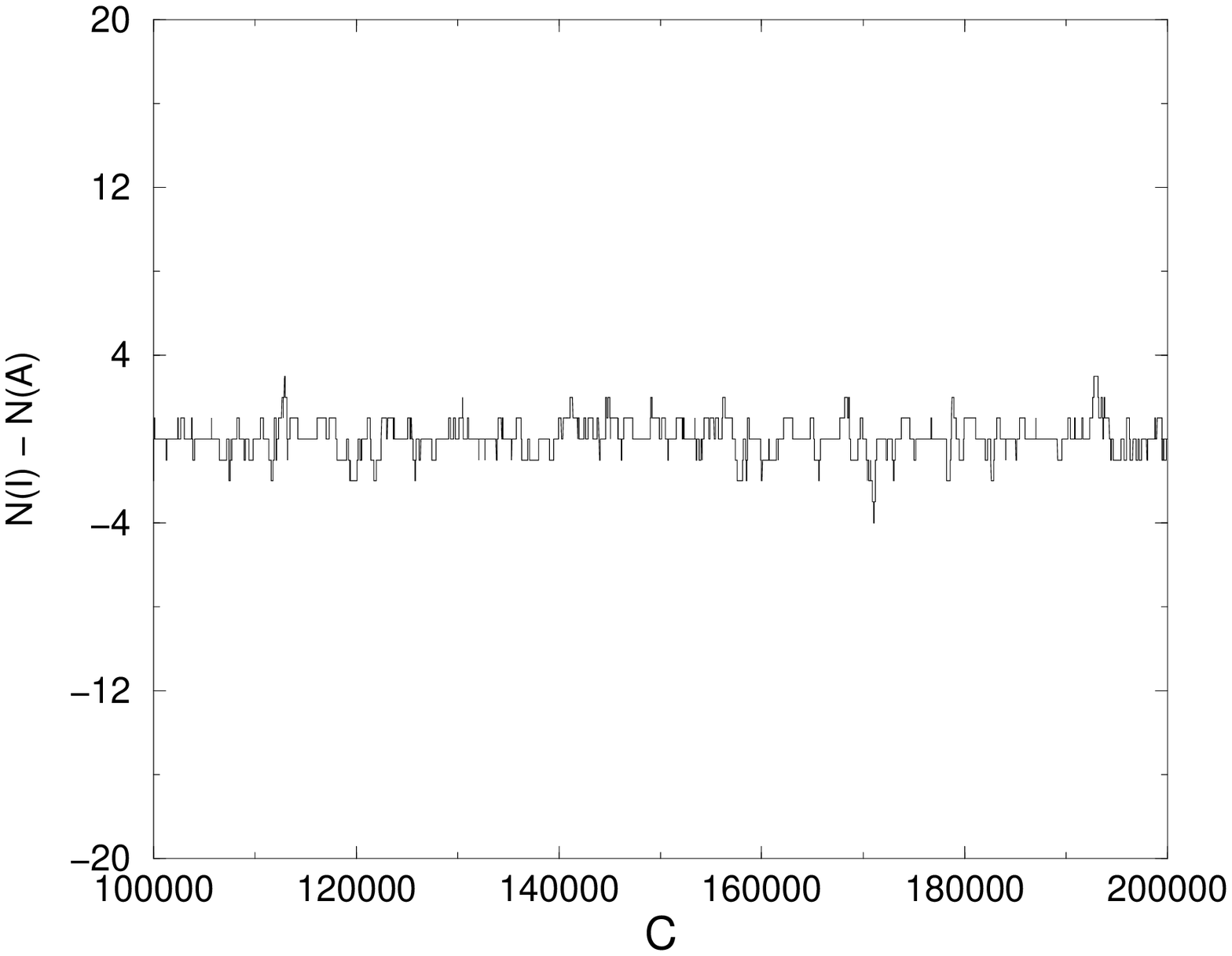}
\end{center}
\caption{$N_{f}=2,\ \overline{\lambda}_{NZ}=2.0,\ V=1,\ m = 0.15$. The
net winding number as a function of the configuration number. We show
only a section of the run for clarity.}
\label{fig:015_eqw_rm2}
\end{figure}

Before proceeding, a note of warning. It is of course trivial 
to vary the number of fermion flavours: all that is required 
is to take the ratio of the determinants to the power $N_{f}$ 
in the Metropolis step. Not surprisingly, however, this
makes the exploration of the available phase space much slower.
This is illustrated in figure~\ref{fig:015_eqt_rm2} where we show 
the total number of topological objects in a Monte Carlo generated
sequence of configurations at our lowest mass value, $m=0.15$.
The corresponding plot for $N_f = 1$ was shown in  
figure~\ref{fig:015_eqt_rm1}. The variation of $N$ is very much
slower in the case of 2 flavours. It is clear that if we want to
go to smaller masses or larger values of $N_f$ we shall need
either much longer Monte Carlo sequences or improved algorithms.
This point is further emphasised by figure~\ref{fig:015_eqw_rm2} 
which displays the increasing difficulty in moving between sectors 
of different net topological charge.

\begin{figure}[tb]
\begin{center}
\leavevmode
\epsfxsize=100mm
\epsfbox{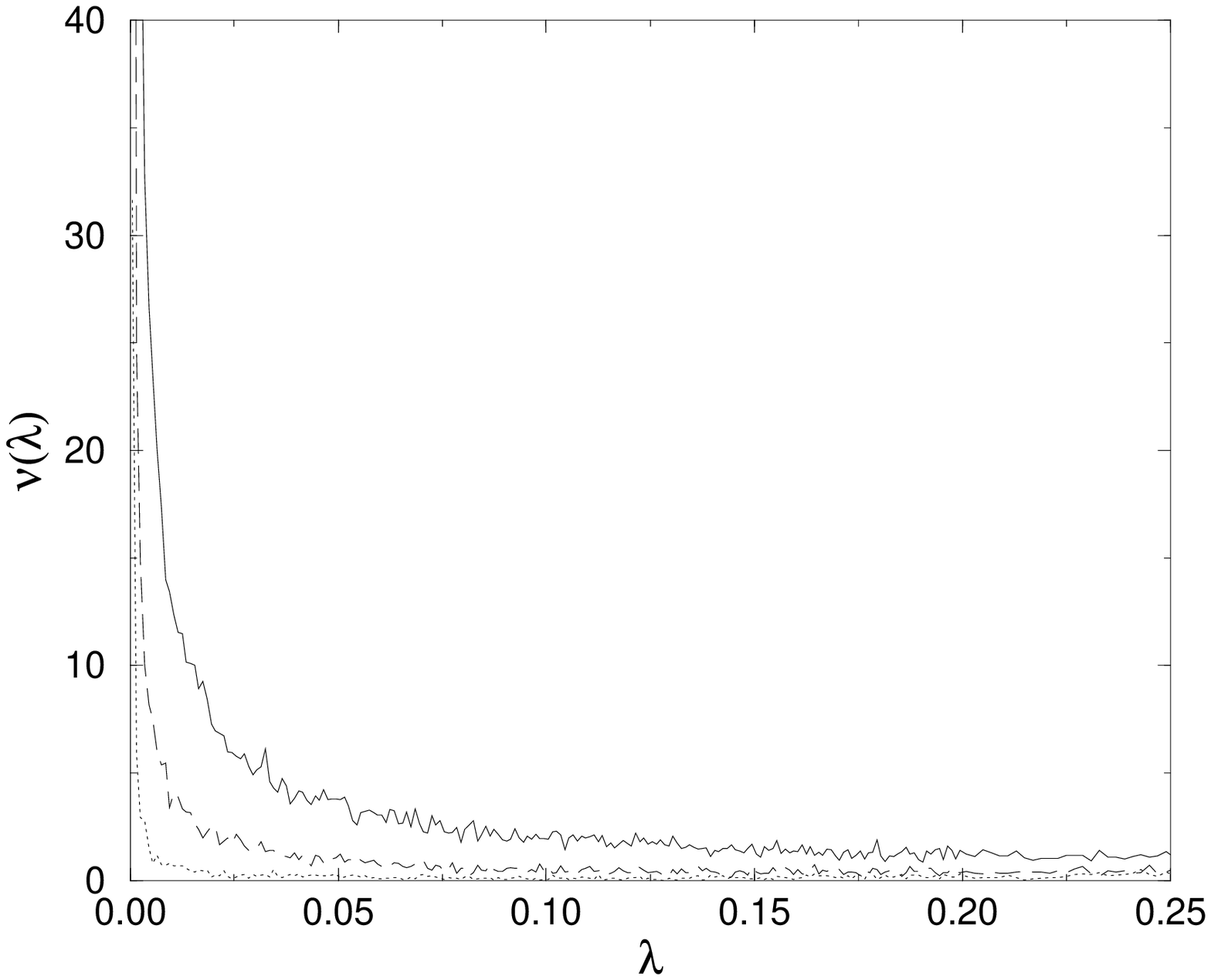}
\end{center}
\caption{$N_{f}=2,\ \overline{\lambda}_{NZ}=2.0,\ V=1$. The spectral
density for quark mass $m=0.5$ (solid), 0.3 (long dashed) and 0.15
(dotted).}
\label{fig:sp_un_rm2}
\end{figure}

\begin{figure}[tb]
\begin{center}
\leavevmode
\epsfxsize=100mm
\epsfbox{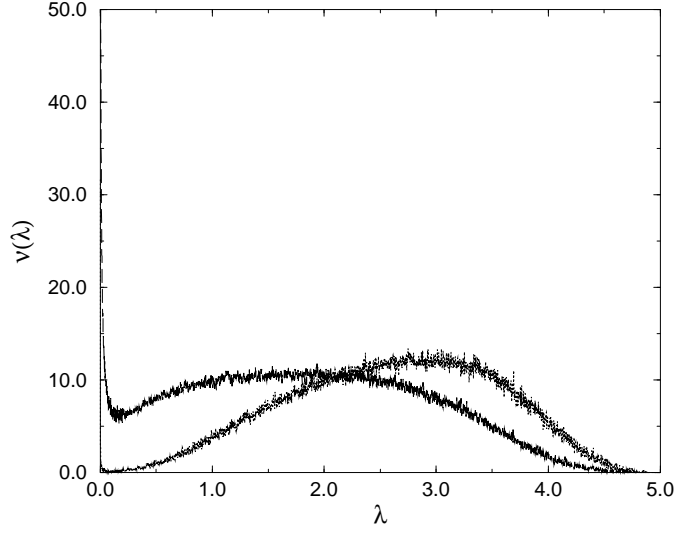}
\end{center}
\caption{$\overline{\lambda}_{NZ}=2.0,\ V=1,\ m=0.15$. A comparison of
the spectral densities obtained for $N_{f}=1$ (dark-solid) and
$N_{f}=2$ (light-dotted).}
\label{fig:sp_un_rm12_cmp}
\end{figure}

\begin{figure}[tb]
\begin{center}
\leavevmode
\epsfxsize=100mm
\epsfbox{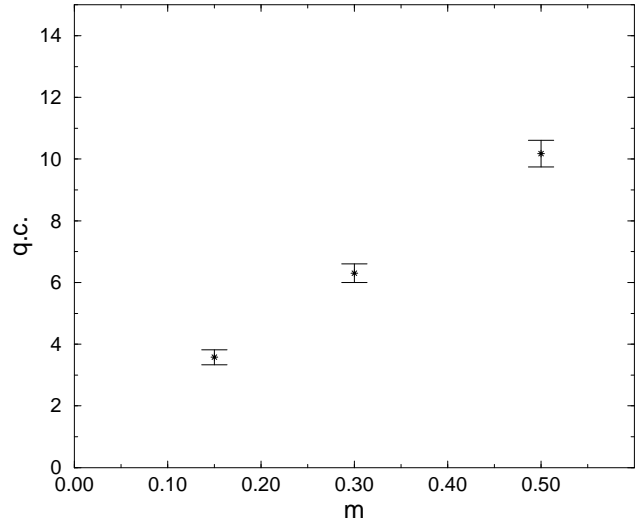}
\end{center}
\caption{The quark condensate \ssi\ as a function of the quark mass
$m$ from the $N_{f}=2$ spectra plotted in figure~\ref{fig:sp_un_rm2}.}
\label{fig:qc_rm2}
\end{figure}

The  $N_f=2$ calculations with variable $N$ have been 
performed for masses $m = 0.15,\ 0.3,\ 0.5$ (chosen,
as before, so that $m \ll \overline{\lambda}_{NZ}$).
The spectral densities for these three masses are depicted in
figure~\ref{fig:sp_un_rm2}. A quick comparison with
figure~\ref{fig:sp_un_rm1} indicates that something very different 
is occuring for two flavours of fermions. The spectral density is
much smaller in magnitude, and decreasing rapidly as $m \to 0$:
all the signs of  chiral symmetry restoration. This is
highlighted by a direct comparison of the $N_{f}=1$ and $N_{f}=2$
spectra at  $m=0.15$,  as shown in figure~\ref{fig:sp_un_rm12_cmp}.
And indeed, when we calculate the chiral condensate from
the $N_f = 2$ spectral densities, we obtain a 
condensate, as shown in figure~\ref{fig:qc_rm2}, that is quite
clearly vanishing as $m \to 0$. 

\begin{figure}[tb]
\begin{center}
\leavevmode
\epsfxsize=100mm
\epsfbox{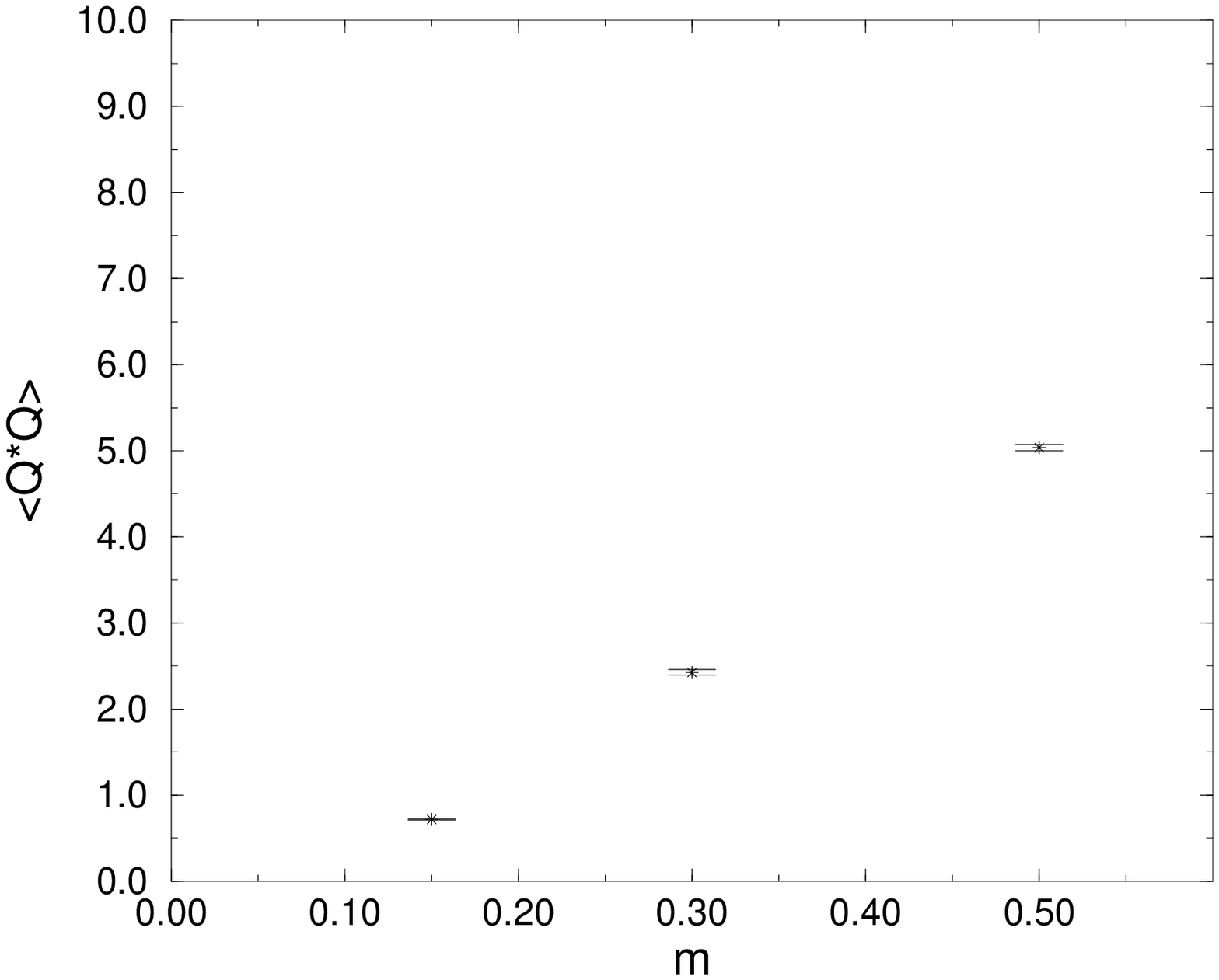}
\end{center}
\caption{$N_{f}=2,\ \overline{\lambda}_{NZ}=2.0,\ V=1$. The second
moment of the winding number distribution $\langle Q^{2}\rangle$ as a
function of the quark mass $m$.}
\label{fig:qq_rm2}
\end{figure}

We also calculate $\langle Q^{2}\rangle$, as listed in 
table~\ref{tab:unq}. If we plot the values versus $m$, 
as in figure~\ref{fig:qq_rm2}, we see that the dependence
is not linear, as one would expect in the phase with symmetry 
breaking, but rather is roughly $\propto m^{N_f} = m^2$ as 
one would expect in the chirally symmetric phase.

Our calculations are thus consistent with the theory being
in a phase with explicit chiral symmetry. Now, while
one does indeed expect to find chiral symmetry restoration 
as we increase $N_f$, it is not expected to occur before 
$N_f = 6$ or so \cite{nf_large}. So if it already happens 
in our calculations for $N_f=2$ this would suggest that our
model provides a rather poor representation of the fermionic
physics in QCD.

There is another possibility: that for $m \leq 0.3$ we have
been driven by finite-volume effects into a symmetry
restored phase. A mass of $m = 0.3$ corresponds to 
$m \sim 25 MeV$ (using our usual criterion for converting 
to physical units) and this is surely small enough, in a 
volume $V \sim 2.5fm^4$, to lead to strong finite size effects. 
Moreover, such a change of phase has been observed in lattice 
QCD simulations \cite{kogut_SU2}. To explore this
possibility we need to perform calculations for larger $m$.
Since we are reluctant to use $m \sim \overline{\lambda}_{NZ}$,
we now turn to the fixed-$N$ calculations.

\subsubsection{$N_{f} = 2$; fixed $N$.}

\begin{figure}[tb]
\begin{center}
\leavevmode
\epsfxsize=80mm
\epsfbox{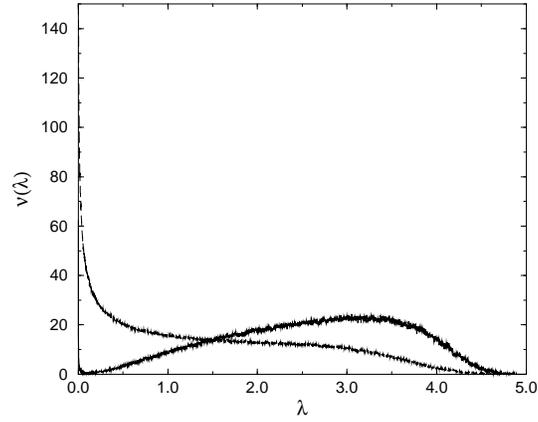}
\end{center}
\caption{$N_{f}=2$, Fixed $N,\ V=1$. The spectral density for
quark mass $m=0.15$ (solid) and $m=3.0$ (long dashed).}
\label{fig:sp_un_rx2}
\end{figure}

Our fixed $N$ calculations are performed over the range
$m \in [0.15,3.0]$. In the mass range $m \leq 0.50$
we find the  same qualitative behaviour as we
obtained in the variable-$N$ calculations. However,
when we compare the  $m=0.15$ and $m=3.0$ spectra, as
in figure~\ref{fig:sp_un_rx2}, we immediately see a great
difference between the two: the spectra corresponding to 
the smaller mass shows the depletion of eigenvalues at small 
$\lambda$ that we expect to see if chiral symmetry is to be 
restored (compare with figure~\ref{fig:sp_un_rm12_cmp}); 
the spectrum corresponding to $m=3.0$ shows no such depletion.

\begin{figure}[tb]
\begin{center}
\leavevmode
\epsfxsize=80mm
\epsfbox{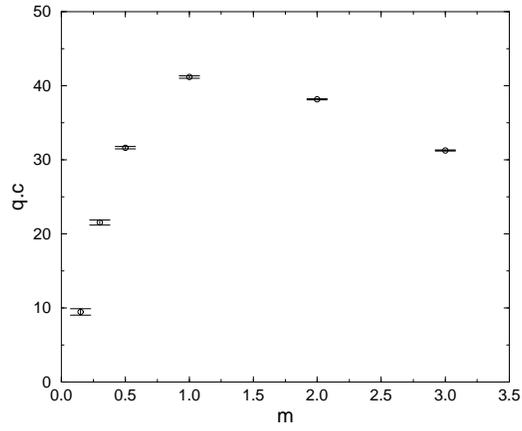}
\end{center}
\caption{$N_{f}=2$, Fixed $N,\ V=1$. The quark condensate \ssi\ as
a function of the quark mass $m$.}
\label{fig:qc_rx2}
\end{figure}

\begin{figure}[tb]
\begin{center}
\leavevmode
\epsfxsize=80mm
\epsfbox{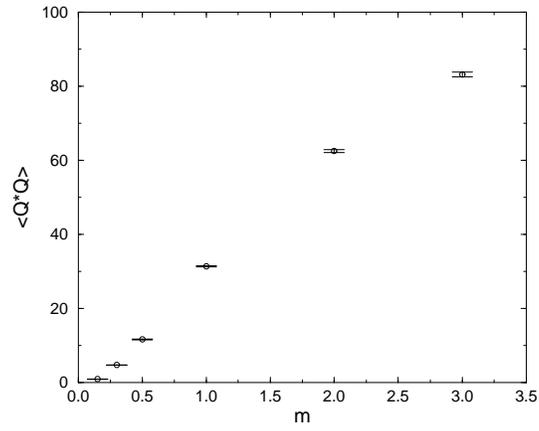}
\end{center}
\caption{$N_{f}=2$, Fixed $N,\ V=1$.  The second moment of the
winding number distribution $\langle Q^{2}\rangle$ as a function of
the quark mass $m$.}
\label{fig:qq_rx2}
\end{figure}

We integrate the spectra for the various masses, and plot the
resulting chiral condensate in figure~\ref{fig:qc_rx2}.
This figure appears to confirm our speculation
that chiral symmetry restoration is a small-$m$ finite
volume effect. For $m \geq 1.0 \sim 80MeV$ we have a chiral
condensate that appears to be heading towards a non-zero
$m = 0$ intercept. For $m < 1.0$ there is a sudden and drastic 
suppression. Such a behaviour is what one would expect 
if the restoration of chiral symmetry were due to finite volume 
effects. The corresponding plot of $\langle Q^{2}(m) \rangle$ is 
given in figure~\ref{fig:qq_rx2}. We see an approximately linear
behaviour for the larger quark masses, as we would expect for 
chiral symmetry breakdown, and, approximately quadratic behaviour 
for the smaller quark masses as before (see figure~\ref{fig:qq_rm2}).
All this strongly suggests that instantons do indeed
break chiral symmetry for two flavours, but that finite
volume effects set in earlier than with one flavour. Of course, 
one really needs to perform such calculations on larger volumes, 
in order to render this plausible scenario completely convincing.  

We note that all the $N_f =2$ spectral densities have a
power-like divergence at $\lambda = 0$. The parameters of the 
fits to $b/\lambda^d$ are listed in table~\ref{tab:sp}. For 
those values of $m$ where we have chiral symmetry breaking the
power of the divergence appears to be $d \simeq 0.3$ as
for $N_f=1$. As we enter the range of $m$ where chiral
symmetry is restored, the value of $d$ increases, and the
coefficient $b$ decreases very rapidly. (Note that this 
supports our earlier suggestion that our $N_f = 1$ calculation
with $m = 0.15$ is afflicted by similar finite-$V$ effects.) 
There is some evidence from the $m = 1.0$ and $2.0$ fits that
$b(m)/\lambda^{d(m)} = b_0 (m/\lambda)^{d_0}$ as for $N_f =1$.

Our estimates for the $\eta^{'}$ mass are listed in
table~\ref{tab:sp}. We see that for values of $m$ where
we observe the spontaneous breaking of chiral symmetry, the
mass is very similar for $N_f = 1$ and $N_f = 2$. The
restoration of chiral symmetry is associated with
$m_{\eta^{'}}$ becoming much heavier (and, indeed, much
harder to estimate reliably).

\section{Conclusions}
\label{section-conclusions}

We have used a simple model for the way instanton zero-modes mix
with each other, in order to learn how topology contributes to 
the Dirac spectral density, particularly at the small eigenvalues 
that are relevant to chiral symmetry breaking. We have done so 
for quenched QCD and for full QCD with both one and two flavours.  

We have found that instantons do indeed appear to break chiral 
symmetry spontaneously, both in quenched QCD and in full QCD. 
We also found that the topological susceptibility appears to 
vanish linearly with the quark mass $m$, as required by the QCD 
anomalous Ward identities when in a phase with chiral symmetry 
breaking. It is remarkable that our simple model manages to 
capture the subtle trade-off between exact zero-modes and 
the chiral symmetry breaking modes arbitrarily close to zero, 
which underlies this result.

That instantons should contribute to chiral symmetry breaking 
is not unexpected\cite{CDG-Carlitz,Diak,Shuryak}.
Our second result is less conventional. We find that the
Dirac spectral density one obtains from instantons has a
powerlike divergence as $\lambda \to 0$. In quenched QCD this
$\sim 1/\lambda^d$ behaviour translates into a pathological
$1/m^d$ divergence in the chiral condensate. In the case of
full QCD the divergence is pushed to smaller eigenvalues as
$\lambda \to 0$, perhaps having the form $(m/\lambda)^d$, 
and the chiral condensate is well-behaved. Nonetheless
such a divergence means that the $\lambda \to 0$ and
$m \to 0$ limits need special care, thus undermining, for 
example, the usual Banks-Casher relation. 

These conclusions broadly confirm the results of earlier work
\cite{ndmt,ndmt2} which used (essentially) the same model. The 
most important way in which we have extended that work, at least 
for quenched QCD, is to show, by varying the various components 
of the model, that the divergence appears to be a generic feature 
of instanton mixing.  In addition, we have also shown that the 
strength of the divergence depends on the instanton density,
becoming negligible for very high densities. In fact it seems 
that `realistic' instanton ensembles, as produced via lattice 
simulations\cite{DSMT}, fall precisely in the intermediate range 
of densities where the strength of the divergence is varying 
most rapidly\cite{usmt2}. Thus we are not able to predict 
the precise phenomenological impact of this divergence upon 
quenched QCD. However it is interesting that when we calculate
the instanton mixing within these lattice ensembles, we find
that the smaller instantons contribute disproportionately
to the smallest eigenvalues; as though they were part of a
more dilute gas. This may provide the crucial link between 
the rather dense gases of large instantons that many lattice 
studies obtain, and the much more dilute gases of significantly
smaller instantons that appear to be required for the 
phenomenological success of liquid instanton models\cite{Shuryak}. 

For full QCD the range of our calculations has been necessarily 
limited and  further work is needed. We have performed
exploratory calculations of some flavour-singlet meson masses
and appear to obtain an $\eta^{\prime}$ mass that remains
massive in the chiral limit, and is $\sim 1 GeV$
if we introduce `physical units' in the simplest possible
fashion. Mass calculations using the eigenvectors of the
Dirac operator would be of particular interest, not least
because they would allow us to test for the presence of a
massless Goldstone pion. Moreover, the fact that the eigenvectors 
become much more extended in space-time as $\lambda \to 0$ 
(as we saw in quenched QCD) should have phenomenological 
consequences.

While we now feel that we have convincing evidence for the
claim that the mixing of would-be instanton zero-modes will 
generically produce chiral symmetry breaking and a 
$\lambda \to 0$ divergence in the Dirac spectral density, 
we cannot of course be sure
that this survives in the full field theory. For example,
if there is some other dynamical mechanism that produces 
a non-zero density of modes at $\lambda = 0$, then the
latter will mix with the (mixed) would-be zero-modes near
$\lambda = 0$, and we cannot be certain that the 
divergence in the latter will survive. 
Equally, if it should be that the modes in the divergent
peak have a space-time extent that $\to \infty$ as
$\lambda \to 0$, then they might well be suppressed by
the confining fluctuations of the theory. There are obvious ways 
that we can, and should, investigate these possibilities
within our model.

An intriguing question is whether there is any relation 
between the divergence that we have found and the logarthmic 
divergence that appears in quenched chiral perturbation 
theory\cite{cpt-log}. A link is plausible because the strength of 
the chiral logarithm is determined by the strength, $\delta$, 
of the pseudoscalar flavour singlet annihilation diagram
which itself is determined by the topological fluctuations
of the quenched vacuum (and whose iteration, in full QCD, 
produces the $\eta^{\prime}$ mass). If $\delta$
is indeed small at the large instanton densities where 
our divergence becomes weak enough to be fit with a logarithm
\cite{usmt2}, then the consistency of the two approaches 
becomes possible. Indeed one can argue\cite{cpt-power} that 
the chiral logarithm will become a power if one sums leading
logs to all orders in chiral perturbation theory. Of course 
once $\delta$ becomes large these perturbative calculations
might become unreliable. In that case our model may provide 
a non-perturbative method for calculating the fate of these 
quenched chiral logarithms. Of course, one might ask how
the power divergence that we see in full QCD fits into all this.
We conjecture that it might be related, in a fashion analogous
to the above, to the remnant chiral logarithms in partially 
quenched QCD. This scenario can and should be tested within 
our model.

\section{Acknowledgements}
MT wishes to acknowledge Nigel Dowrick's important influence on this
work, through their extensive, but largely unpublished, collaboration
some seven years ago. We
wish to thank PPARC for support under grant GR/K55752. US also wishes
to thank PPARC for a research studentship (number 96314624). The
computations were performed on our Departmental workstations.

\begin{table}[htb]
\begin{center}
\begin{tabular}{|c|c|c|c|c|c|c|}
Type & $f$ & $V$ & $N_{c}$ & d & $\chi_{p}^{2}/N_{DF}$ &
$\chi_{l}^{2}/N_{DF}$\\
\hline
Hard Sphere & 0.2 & 1.0 & 650000 & 0.656$\pm$0.003 & 2.40 & 600\\
& 0.5 & 1.0 & 128000 & 0.695$\pm$0.002 &  1.92 & 508\\
& 1.0 & 0.41 & 130000 & 0.540$\pm$0.003 &  1.71 & 123\\
& 1.0 & 1.0 & 126000 & 0.595$\pm$0.002 & 1.38 & 258\\
& 1.0 & 1.52 & 97000 & 0.617$\pm$0.001 & 1.23 & 400\\
& 1.0 & 2.44 & 93000 & 0.640$\pm$0.001 & 1.47 & 702\\
& 1.0 & 7.72 & 9780 & 0.668$\pm$0.003 & 1.73 & 456\\
& 1.75 & 1.0 & 111000 & 0.309$\pm$0.002 & 1.94 & 48\\
& 2.5 & 1.0 & 63200 & 0.075$\pm$0.014 & 1.72 & 2.59\\
& 5.3 & 1.0 & 6720 &  0.004$\pm$0.010 & 2.20 & 2.19\\
& 10.0 & 1.0 & 1266 & 0.008$\pm$0.016 & 1.91 & 1.90\\
Gaussian & 1.0 & 1.0 & 12600 & 0.588$\pm$0.005 & 1.36 & 45\\
& 1.0 & 2.4 & 15500 & 0.634$\pm$0.002 & 1.31 & 201\\
& 2.5 & 1.0 & 6320 & 0.290$\pm$0.007 & 1.46 & 4.87\\
Classical & 1.0 & 1.0 & 2100 & 0.477$\pm$0.010 & 1.32 & 5.33\\
& 1.0 & 2.4 & 450 & 0.538$\pm$0.022 & 1.38 & 3.50\\
Class. (II) & 1.0 & 1.0 & 2520 & 0.555$\pm$0.006 & 1.56 & 34.6\\
\end{tabular}
\end{center}
\caption{Parameters and results for $Q=0$, $N_f=0$ ensembles. 
$f$ is the packing fraction, $V$ is the space-time volume,
$N_{c}$ is the number of configurations in the
ensemble, $d$ is the degree of divergence, $\chi_{p}^{2}/N_{DF}$ 
is the chi-square for the best power fit to the divergence and 
$\chi_{l}^{2}/N_{DF}$ is the chi-square for the best log fit.}
\label{tab:basicdat}
\end{table}

\begin{table}[htb]
\begin{center}
\begin{tabular}{|c|c|c|c|c|c|c|}
Type & $f$ & $V$ & $N_{c}$ & d & $\chi_{p}^{2}/N_{DF}$ & $\chi_{l}^{2}/N_{DF}$\\ \hline
Hard Sphere & 1.0 & 1.0 & 126000 & 0.363$\pm$0.006 & 1.90 & 65\\
& 1.0 & 2.4 & 62000 & 0.522$\pm$0.003 & 2.07 & 253\\
\end{tabular}
\end{center}
\caption{Parameters and results for $N_f=0$ ensembles with variable
$Q$ and $N$. $f$ is now the mean packing fraction.}
\label{tab:varyq}
\end{table}

\begin{table}[htb]
\begin{center}
\begin{tabular}{|c|c|c|c|c|c|}
Type & $f$ & $V$ & $\delta\rho(0,2)$ & $\delta\rho(2,4)$ & $\delta\rho(4,6)$\\
\hline
Hard Sphere & 0.77 & 0.41 & 0.88(3) & 0.75(4) & 0.68(3)\\
& 2.96 & 0.11 & 0.85(3) & 0.75(3) & 0.67(3)
\end{tabular}
\end{center}
\caption{Values of $\delta\rho(|Q|,|Q|+2)$, as defined in the text,
for two different $N_f=0$ hard sphere ensembles.}
\label{try_ndmt}
\end{table}

\begin{table}[tbh]
\begin{center}
\begin{tabular}{|c|c|c|l|l|c|}
\hline
Set & $\overline{\lambda}_{NZ}$ & $N_{f}$ & $m$ & 
$\langle Q^{2}\rangle$ & $\langle N_{I}\rangle$\\
\hline
A & 2.0 & 1 & 0.5 & 25.65$\pm$0.45 & 35.53\\
B & 2.0 & 1 & 0.3 & 14.28$\pm$0.16 & 34.91\\
C & 2.0 & 1 & 0.15 & 8.13$\pm$0.04 & 35.75\\
D & 2.0 & 2 & 0.5 & 5.04$\pm$0.04 & 21.35\\
E & 2.0 & 2 & 0.3 & 2.43$\pm$0.03 & 23.17\\
F & 2.0 & 2 & 0.15 & 0.72$\pm$0.01 & 30.85\\
G & - & 1 & 3.0 & 98.615$\pm$0.78 & 62.80\\
H & - & 1 & 2.0 & 80.097$\pm$0.92 & 63.03\\
I & - & 1 & 1.0 & 60.383$\pm$0.49 & 62.95\\
J & - & 1 & 0.5 & 34.22$\pm$0.26 & 62.98\\
K & - & 1 & 0.3 & 22.00$\pm$0.19 & 62.99\\
L & - & 1 & 0.15 & 11.66$\pm$0.07 & 63.01\\
M & - & 2 & 3.0 & 83.20$\pm$0.64 & 63.07\\
N & - & 2 & 2.0 & 62.49$\pm$0.40 & 62.90\\
O & - & 2 & 1.0 & 31.37$\pm$0.13 & 62.93\\
P & - & 2 & 0.5 & 11.61$\pm$0.10 & 62.98\\
Q & - & 2 & 0.3 & 4.71$\pm$0.05 & 62.96\\
R & - & 2 & 0.15 & 0.89$\pm$0.02 & 62.98\\
\hline
\end{tabular}
\end{center}
\caption{Some information about the $N_f\not=0$ full QCD ensembles.
Where $N$ has been kept fixed, a value for $\overline{\lambda}_{NZ}$
is not needed, and is not given.}
\label{tab:unq}
\end{table}

\begin{table}[tbh]
\begin{center}
\begin{tabular}{|c|c|c|c||c|c|}
\hline
Set & b & d & $\chi^{2}/N_{DF}$ & $m_{\eta^{'}}$ & $m_{\sigma}$ \\
\hline
A & 2.904$\pm$0.158 & 0.668$\pm$0.008 & 1.09 
& 11.2$\pm$2.0 & 11.3$\pm$2.5 \\
B & 1.956$\pm$0.181 & 0.699$\pm$0.017 & 0.94 
& 12.9$\pm$2.5 & 10.9$\pm$1.8 \\
C & 0.491$\pm$0.058 & 0.861$\pm$0.022 & 2.41 
& 11.5$\pm$1.5 & 10.2$\pm$2.4 \\
D & 0.228$\pm$0.031 & 0.882$\pm$0.021 & 1.15 
& 17.3$\pm$1.9 & 15.2$\pm$2.0 \\
E & 0.043$\pm$0.009 & 0.979$\pm$0.032 & 1.44 
& 19.3$\pm$2.0 & 17.4$\pm$2.7 \\
F & - & - & - 
& 20.0$\pm$2.0 & 16.7$\pm$2.2 \\
G & 30.917$\pm$3.324 & 0.300$\pm$0.018 & 0.99 
& 8.9$\pm$3.1 & - \\
H & 28.766$\pm$2.058 & 0.307$\pm$0.013 & 1.35  
& 8.5$\pm$2.5 & - \\
I & 36.517$\pm$4.636 & 0.252$\pm$0.019 & 1.41 
& 11.5$\pm$2.2 & - \\
J & 24.048$\pm$2.936 & 0.288$\pm$0.019 & 2.05 
& 11.7$\pm$2.1 & - \\
K & 17.599$\pm$2.632 & 0.309$\pm$0.023 & 3.27 
& 12.4$\pm$2.9 & - \\
L & 4.395$\pm$0.586 & 0.478$\pm$0.024 & 4.20  
& 12.1$\pm$2.3 & - \\
M & 30.410$\pm$3.919 & 0.296$\pm$0.020 & 1.17 
& 9.5$\pm$3.7 & - \\
N & 30.990$\pm$2.520 & 0.285$\pm$0.014 & 1.57 
& 13.6$\pm$2.7 & - \\
O & 17.665$\pm$1.714 & 0.343$\pm$0.017 & 1.41 
& 15.0$\pm$3.6 & - \\
P & 4.404$\pm$0.267 & 0.519$\pm$0.013 & 1.99 
& 20.6$\pm$2.5 & - \\
Q & 0.580$\pm$0.051 & 0.788$\pm$0.019 & 1.55 
& 15.8$\pm$2.5 & - \\
R & 0.006$\pm$0.002 & 1.307$\pm$0.043 & 2.30 
& 18.2$\pm$2.1 & - \\
\hline
\end{tabular}
\end{center}
\caption{Parameters of power-law fits to the spectral densities, 
for the ensembles given in table~\ref{tab:unq}, as well as
mass estimates for the  $\eta^{'}$ and  $\sigma$ mesons.}
\label{tab:sp}
\end{table}

%\begin{figure}[tb]
%\begin{center}
%\leavevmode
%\epsfxsize=100mm
%\epsfbox{mass_eta_sig.eps}
%\end{center}
%\caption{$N_{f}=1,\ V=1$. The mass of the  $\eta^{'}$ as a function 
%of the quark mass.}
%\label{eta_sig_mass}
%\end{figure}

\end{document}